\documentclass[aps, nofootinbib]{revtex4}
\usepackage{amssymb}
\usepackage{amsmath}
\usepackage{graphicx}
\usepackage{psfrag}
\newcommand{\be}{\begin{equation}}
\newcommand{\ee}{\end{equation}}
\newcommand{\bes}{\begin{eqnarray}}
\newcommand{\ees}{\end{eqnarray}}
\newcommand{\R}{\mathbb{R}}
\newcommand{\C}{\mathbb{C}}

\newcommand{\SO}{\mathrm{SO}}
\newcommand{\SU}{\mathrm{SU}}
\newcommand{\U}{\mathrm{U}}

\newcommand{\su}{\mathfrak{su}}

\renewcommand{\det}{\mathrm{det}}
\newcommand{\tr}{\mathrm{tr}}

\newcommand{\pd}{\partial}
\newcommand{\rar}{\rightarrow}
\newcommand{\f}{\frac}

\newcommand{\cD}{{\cal D}}
\newcommand{\cH}{{\cal H}}
\newcommand{\cM}{{\cal M}}
\newcommand{\cO}{{\cal O}}
\newcommand{\cK}{{\cal K}}
\newcommand{\cP}{{\cal P}}
\newcommand{\cV}{{\cal V}}
\newcommand{\cS}{{\cal S}}
\newcommand{\cZ}{{\cal Z}}

\newcommand{\dman}{\Delta}
\newcommand{\dmans}{\Delta^*}

\newcommand{\bE}{\bar{E}}
\newcommand{\bA}{\bar{A}}

\newcommand{\vs}{v^*}
\newcommand{\es}{e^*}
\newcommand{\fs}{f^*}

\newcommand{\vb}{\bar{v}}
\newcommand{\eb}{\bar{e}}

\newcommand{\vbs}{\bar{v}^*}
\newcommand{\ebs}{\bar{e}^*}

\newcommand{\Ts}{T^*}

\newcommand{\gal}{\alpha}
\newcommand{\gde}{\delta}
\newcommand{\gDe}{\Delta}
\newcommand{\gep}{\epsilon}
\newcommand{\gka}{\kappa}
\newcommand{\gla}{\lambda}
\newcommand{\gLa}{\Lambda}

\newcommand{\gga}{\gamma}
\newcommand{\gGa}{\Gamma}
\newcommand{\gsi}{\sigma}
\newcommand{\gSi}{\Sigma}
\newcommand{\gth}{\theta}

\newcommand{\field}[3]{\phi(g_{#1}, g_{#2}, g_{#3})}

\begin{document}
%
%
\begin{titlepage}
\title{\large A new proposal for group field theory I: the 3d case}
\author{ James Ryan}
\email{j.p.ryan@damtp.cam.ac.uk} 
\affiliation{\vspace{2mm} Department of Applied Mathematics and Theoretical Physics,\\ 
Centre for Mathematical Sciences,\\ Wilberforce Road,\\ Cambridge CB3 0WA, UK}
\date{\today}
\begin{abstract}
\begin{center}{\bf Abstract }\end{center}
In this series of papers, we propose a new rendition of $3d$ and $4d$ state sum models based upon the group field theory (GFT) approach to non-perturbative quantum gravity.  We will see that the group field theories investigated in the literature to date are, when judged from the position of quantum field theory, an unusual manifestation of quantum dynamics.  They are one in which the Hadamard function for the field theory propagates a-causally the physical degrees of freedom of quantum gravity.  This is fine if we wish to define a scalar product on the physical state space, but it is not what we generally think of as originating directly from a field theory.  We propose a model in $3d$ more in line with standard quantum field theory, and therefore the field theory precipitates causal dynamics.  Thereafter, we couple the model to point matter, and extract from the GFT the effective non-commutative field theory describing the matter dynamics on a quantum gravity background.  We identify the symmetries of our new model and clarify their meaning in the GFT setting.  We are aided in this process by identifying the category theory foundations of this GFT which, moreover, propel us towards a categorified version for the $4d$ case.
\end{abstract}
\maketitle
\end{titlepage}
%
\section{Introduction}
Currently, spin foam models \cite{Ori_review, Per_review} draw considerable interest as a general formalism for quantum gravity.  The reasons are multifarious.  For instance, they occur at a point of convergence of different objectives, including loop quantum gravity \cite{Ash_lectures, RovSmo_loop}, topological field theories \cite{Topological}, and simplicial gravity.  Indeed, the spin foam picture emerges when considering the evolution in time of spin networks - kinematical states of quantum general relativity (ensuing from loop quantum gravity).  As we said, it arises in the development of topological field theories including $3d$ gravity.  In these models, category theory plays a major role, since their whole construction can be rephrased in terms of operations in the category of Lie groups.

On a different tack, spin foams naturally arise as lattice discretizations of the path integral for gravity and generally covariant gauge theories.  Moreover, they provide a background independent discrete quantum gravity path integral, by representing space-time as a combinatorial cellular complex upon which we encode geometric data in a purely group-theoretic and algebraic manner.  This encoding may be done in two equivalent ways: in the \lq configuration-space' representation where the geometry labels for the cellular 2-complex are Lorentz group representations; or alternatively, in the \lq momentum-space' representation where the labels occur as Lorentz group elements.  A colouring consistent with a choice of geometry is an admissible configuration, and once this has been completed, we proceed to assign it a quantum amplitude. The discretization reduces the number of degrees of freedom to finitely many, thus enabling the definition of a functional measure.  This makes the formulation seem similar to standard lattice gauge theory.  But although the manner of the truncations are akin to each other, its nature in the quantum gravity setting is intrinsically different to that of conventional lattice theory.  As expected, background independence tells us that we cannot consider the cellular complex as a UV regulator. Summing over admissible configurations weighted by their chosen quantum amplitude gives us our partition function.    The cell complex is usually chosen to be topologically dual to a simplicial complex of the appropriate dimension.

\medskip

In turn, spin foam models have been obtained from so-called group field theories \cite{Ori_gftreview, Fre_gftreview}.  A group field theory (GFT), as its name suggests, is a field theory over a group manifold, which generates a sum over all cellular 2-complexes in its Feynman diagram expansion.  An equivalent way of expressing this result is to say that we obtain a sum over all coloured simplicial complexes, by the duality mentioned above, and thus a sum over all geometries and topologies, of a given dimension.  The impact of such a formalism is two-fold.  Unlike gravity in 3d, the 4-dimensional theory is not a topological field theory and thus contains more than just global degrees of freedom. Indeed, we can read in any number of expositions about the infinite set of local degrees of freedom inherent in any theory based on 4d gravity.  Thus, inserting a fixed, albeit coloured, discrete structure such as a cellular complex, destroys the background independence of our resultant quantum theory.  But this can be restored by a summation over geometries, that is, a summation over all simplicial complexes discretizing a given manifold $\cM$.  The group field theory provides a well defined prescription for implementing this sum.  But it goes further, in that there is a sum over manifolds $\cM$ in the GFT, thus realizing a dream of many relativists to make yet another of the fundamental structures of nature dynamical, the space-time topology. The quantum dynamics of the GFT can be viewed as a \lq local simplicial third quantization'  of gravity\footnote{Third quantization is a label often used to describe the second quantization of geometry, since the classical variables for gravity are the metric {\it field}.}.  To explicate this point more thoroughly, the GFT filed represents a fundamental building block of space, a $(d-1)$ dimensional ball. A collection of fields comprises a quantum a state of geometry.  Thus, the GFT path integral describes the quantum dynamics of this quantum state.  The classical equations of motion related to this GFT embody the Hamiltonian constraint for gravity, in filed theory language.  Thus, solutions to the classical equations of motion are quantum states of geometry which are solutions to the Hamiltonian constraint.  In other words, solutions are elements of the physical state space of loop quantum gravity. This is in analogy with matter field theory, where the $2nd$ quantized Klein-Gordon field represents a quantum state of geometry.  The path integral then represents its quantum dynamics.  The classical Klein-Gordon field then represents solutions to the Hamiltonian constraint for the particle, where the momentum is on-shell; and thus represents a state in the physical space of the $1st$ quantized theory.   

\medskip

In 4-dimensions, there exist promising and recently much studied spin foam models, and group field theories, with many interesting features and issues still in need of clarification. On the other hand in $3d$, where gravity is a topological field theory, we have a much more complete story.  We know that its quantization derived using spin foam models is equivalent to those obtained by other approaches, such as loop quantum gravity \cite{NouPer_lqgphysical}, Chern-Simons quantization, 't Hooft's polygon approach \cite{tHo_polygon} etc.  Furthermore, $3d$ gravity raises many of the issues involved in the quantization of gravity such as:  the conceptual problem of time, the problem of the construction of physically relevant observables, the emergence of the semiclassical space-time geometry, the effect of a cosmological constant, the quantum causal nature, the role of diffeomorphisms; the sum over topologies etc. Thus, $3d$ gravity provides an excellent testing ground for theories of quantum gravity.

The coupling of matter fields to quantum gravity in the spin foam framework is of major importance. Matter coupling might provide the best, if not the only, way to attack certain issues that are notoriously difficult in a pure gravity theory.  They provide an avenue to define quantum gravity observables that have a clear physical meaning. Furthermore, they stimulate a program of quantum gravity analysis, supplementing the hypothesis that quantum gravity affects the usual predictions of quantum field theory.  Over and above this, quantum gravity might hopefully solve problems in quantum field theory.  For example, quantum field theories are in general plagued by ultraviolet divergences, but quantum gravity might provide a built-in covariant cut-off at the Planck scale.

Recently this agenda has received much attention.  In \cite{NouPer_lqqparticles}, point matter was coupled to gravity from the canonical perspective.  The resulting quantum dynamics are played out on a simplicial manifold of topology $\gSi\times\R$ ($\gSi$ is a Riemann surface), a particular subspecies of spin foams.  The path integral formulation has concurrently made similar progress \cite{FreLou_PR1, FreLou_PR2}. The covariant discretization generalizes the canonical results in the sense that one may deal with a fixed manifold of arbitrary topology.  Work within the covariant formalism oiled the wheels of progress towards a phenomenological understanding of matter in the quantum gravity setting.  By summing over the gravity degrees of freedom, the matter spin foam theory takes on the character of a Feynman diagram of a non-commutative field theory \cite{FreLiv_PR3, KarLivOriRya_spinning}.  The momentum space realization of this theory has support on the group $\SU(2)$, rather than the corresponding Lie algebra $\su(2) \sim \R^3$, and thus has bounded momentum. Hence we are dealing with another incarnation of group field theory.  Since this effective field theory encodes the sum total of the quantum gravity effects in the matter sector, we should be able to extricate the gravity degrees of freedom and embellish the effective theory so that it may be written as a group field theory in the spin foam sense \cite{FreOriRya_gftscalar, OriRya_gftspinning}.

Another facet of discrete quantum gravity currently the subject of much research is the imposition of causal restrictions \cite{LivOri_causal, OriTla_causal}.  When we pen any path integral formulation, we have several alternatives, depending on our motivation \cite{Tei_qmechanics}. One option is to use the sum-over-histories to project the kinematical states down onto their physical subspace, and provide us with a physical inner product.  This amplitude is real, it does not attribute an incoming or outgoing status to the states.  Thus it tenders an a-causal dynamics.  Another choice within the path integral furnishes us with causal dynamics.  Adhering to a covariant prescription implies blindness towards time-ordering.  Knowledge of space-time orientation, however,  is enough to impose a causal structure \cite{Ori_feynman}.  This has already been extended to the group field theory.  These \lq generalized group field theories' \cite{Ori_gengft} register space-time orientation in their operators,  utilizing a mechanism reminiscent of Feynman's proper time formalism.

\medskip

This article shall be devoted to the proposal and study of a new GFT approach. The motivation for this proposal is that when we contemplate group field theory we get caught in the following dilemma.  On the one hand, we rely on an analogy with quantum field theory to justify our belief in group field theory as the fundamental theory.  The other hand, unfortunately, has something in store for us.  The majority of GFTs to date do not include causality, which does not coincide with conventional quantum field theory.  In the standard QFT setting, the kinetic operator inverts to become the Feynman propagator and hence labels causal amplitudes.  The corresponding a-causal dynamics are accommodated by the Hadamard function, a non-invertible operator on the space of fields.  Generalized GFTs remedy this situation, with causal dynamical operators occurring in both the kinetic and vertex operators. They provide, at the moment, the most fundamental implementation of causality in discrete quantum gravity.  They register a specific record of the orientation of the simplicial building blocks, for both space and space-time, and this is shown to generate orientation dependent spin foam quantum amplitudes.  In \cite{OriTla_causal}, causality is included in a superficially different fashion.  The explicit orientation labels are obscured , but  the quantum amplitude registers orientation dependence.  This suggests that we might be able to include causality in a more direct, albeit possibly less fundamental fashion in the GFT.

Moreover, since the a-causal dynamics of earlier GFTs are included in the vertex term, the matter coupling in \cite{FreOriRya_gftscalar, OriRya_gftspinning} is also included completely in the vertex term.  The generalized GFTs corresponding to these theories will again provide an implementation of causality in terms of explicit orientation dependence of the simplices.  Of course, since this dependence is included into the kinetic as well as vertex terms, it provides an avenue to transfer vital matter information from the vertex to kinetic term.  We require this if the generalized GFTs are to reduce to the effective field theory for matter.  But we are again motivated by these complications to opt for a direct manner of including causal matter and causal gravity information in the GFT kinetic term.  

Our proposal alleviates these difficulties by directly placing a causal dynamical operator in the kinetic term of the GFT.  This further facilitates its subsequent reduction to the correct matter effective field theory.  While symmetries in GFT are notoriously difficult to elucidate, our new formalism clarifies their position in the action, with respect to their position at the level of the spin foam amplitudes.  Spin foam models have a categorical description in terms of morphisms, natural transformations, functors etc.  We make this identification explicit in the GFT locale and briefly mention further work in 4d.

Finally, an important step in making this new GFT a reality required a fundamental shift in how we regard the various discrete structures arising in our quantization of gravity.  We do not directly deal with a simplicial complex as a discretization of our manifold.  Instead we deal with the more general CW-complex.  But from the GFT context we cannot generate an arbitrary cell-complex, so we refine our choice.  We discretize our manifold using a cell-complex whose dual structure is a simplicial complex.  To emphasize our point, the spin foam 2-skeleton is now a simplicial complex.  So the spin foam amplitudes are not in general Ponzano-Regge amplitudes, but are an equally valid state-sum for $3d$ gravity.

\medskip

In the next section, we introduce the formal aspects of the path integral quantization of gravity.  We exploit an analogy with the $2nd$ quantization of point particle theory, to illuminate what we demand of our GFT.  In section \ref{3dprelim}, we provide an elementary description of the Ponzano-Regge model for 3d quantum gravity, along with its incorporation of matter and causality, especially with regard to the effective field theory for matter.  Following that, we describe the creation of these amplitudes as the Feynman diagrams of group field theories, again with a concise account of how they incorporate matter features.  In section \ref{new}, we will begin with a precise rendition of how our new viewpoint with respect to the discrete space-time structures, before proceeding onto its embodiment as a group field theory.  We thereafter develop our model to include matter, and reduce it to the effective field theory.  After analyzing the model's symmetries, we finish our exposition with some concluding statements in section \ref{conclusion}.

\section{The path integral quantization for quantum gravity}
\label{formal}
 An elegant analysis of this topic occurs in \cite{LivOri_causal}, so let us be brief. There is a formal analogy between the covariant $1st$ and $2nd$ quantizations of point particle matter and those of gravity.  Here, we shall only outline the gravity case.

\subsection{Quantum gravity}
\label{formal_qgrav}

We define, formally, the path integral approach to quantum gravity where the transition amplitude is a sum over all 4-geometries interpolating between given boundary 3-geometries.  This sum is weighted by the exponential of ($i$ times) the Einstein Hilbert action for general relativity  and a suitable measure on the space of (diffeomorphism classes of) 4-metrics. Furthermore, there is a possible additional sum over all the possible manifolds, i.e topologies, having the given boundary\footnote{This has particular relevance in the group field theory scenario.}.  The action can be re-expressed in its Hamiltonian formulation for manifolds of topology $\cM\sim\gSi\times\R$. Once we transplant to this context, we will see more clearly how explicit traces of causality appear.
\be
\label{formal_hamiact}
 \cS_{\cM}(h, \pi, N^i, N) = \int_\cM d^3x\, dt \,
 (\pi^{ij}{h}_{ij} - N\cH - N^i\cH_i),
\ee
where the variables are $h_{ij}$, the 3-metric induced on a spacelike slice of the manifold $\cM$, $\pi^{ij}$ is its conjugate momentum, the shift $N^i$, a Lagrange multiplier that enforces the spatial diffeomorphism constraint $\cH_i = 0$, and the lapse $N$ that enforces the Hamiltonian constraint $\cH = 0$.  The Hamiltonian constraint encodes the dynamics of the theory and the symmetry under time diffeomorphisms\footnote{We neglect here in this formal discussion to examine an appropriate gauge-fixing with its accompanying ghost terms \cite{Tei_qmechanics}.}.  We neglect to describe more about the former constraint.  We will be more interested in the integration over the lapse where our choice of range for this variable is crucial. If we choose the range $(-\infty,\,\infty)$, then we have projected onto the physical state space of quantum gravity
\be
\label{formal_physical}
 _{phys}\langle h_1| h_2\rangle_{phys} = \int^{+\infty}_{-\infty} \cD N \int_{g|h_1,h_2} 
 \Big( \prod_x \cD g_{ij}(x) \cD\pi^{ij}(x)\Big) e^{i\cS}.
\ee
But this expression is completely invariant under the reversal of space-time orientation. The reason for this is that the canonical algebra generated by $\cH_i$ and $\cH$ induces a larger symmetry than 4-diffeomorphism invariance.  What is more,  like its counterpart in matter field theory, (\ref{formal_physical}) is the Hadamard function for gravity 
\be
\label{formal_qghada}
G_H(h_1,h_2) =  \,_{phys}\langle h_2| h_1\rangle_{phys} = \,_{kin}\langle h_2\mid ``\gde(\cH)" \mid h_1\rangle_{kin}
\ee
Formally, this is a solution of the Wheeler-DeWitt equation in both arguments and does not register any notion of whether a state is incoming or outgoing. If we want to include causality we must restrict the range of the lapse to $(0, +\infty)$. This gives us the quantity analogous to the Feynman propagator
\be
\label{formal_qgfeyn}
 G_F (h_1,h_2) = \int^{+\infty}_{0} \cD N \int_{g|h_1,h_2} 
 \Big( \prod_x \cD g_{ij}(x) \cD\pi^{ij}(x)\Big) e^{i\cS} 
 = \langle h_2\mid h_1\rangle_{C} = 
 \,_{kin}\langle h_2\mid ``\frac{1}{\cH + i\gep}" \mid h_1\rangle_{kin}
\ee

The restriction to positive lapses is a causality restriction in that it corresponds to imposing that the final 3-geometry lies to the future of the first, and imposes a \lq timeless ordering' between initial and final boundary data.  Furthermore, we have relaxed the classical constraint to allow for purely quantum histories, where quantum dynamics takes place off-shell with respect to the classical constraint - a common feature of quantum theories.  It is necessary for the above reason and to make contact with the Langrangian point of view.  

Now, we pass to the $2nd$ quantized path integral formalism.  This reminds us of earlier work on a quantum field of geometry, i.e. a QFT on superspace \cite{Superspace}, for a given spatial topology. Such a theory would generate in its Feynman expansion a sum over different topologies each corresponding to a possible Feynman graph of the theory and to a possible interaction process for  \lq universesÕ represented by a 3-manifold of the given topology. The spatial topology, in this scenario, is restricted but the quantum amplitude for the for each Feynman graph, corresponding to a particular space-time topology with $n$ boundary components, would be  given by a path integral for gravity on that space-time topology.  But there are formidable conceptual and technical difficulties with proposing a field that encompasses a complete 3-manifold state.  We have an alternative. Unlike the point particle case, a state of quantum geometry has internal structure, and we may choose the quantum field to describe only a local subset of this structure.  Thus, the new field will describe a 3-ball in the 3-manifold and will generate the the full state and its evolution through the appropriate definition of boundary observables and an action.  This is what we try do in the GFT formalism. By analogy, we propose that its action should take the form
\be
\label{formal_wdw}
\cS(\phi) = \int db\,d\bar{b} \,\phi(b) K_F(b,\bar{b}) \phi(\bar{b}), 
\ee
where $b$ is an elementary $3$-ball and $K_F(b, \bar{b}) = \cH(b) \times \gde(b, \bar{b})$. We may then add interaction terms to the action in order to generate more interesting space-times, and topology change.  Because the formalism is local, we are no longer restricted to dealing with a fixed spatial topology.  Furthermore, we will deal with this theory in the simplicial setting, where the elementary $3$-balls are $3$-simplices. 

We note here, that this is not how GFTs are usually formulated.  In general, one has trivial dynamics in the kinetic operator and a-causal dynamics in the vertex operator, where the elementary $3$-simplices of space interact to create elementary $4$-simplices of space-time.  This is perfectly natural when we regard the GFTs as a method of constructing a physical inner product on the state space of quantum gravity.  In this setting, the spin foams describe the evolution of spin networks where $4$-simplices embody Hamiltonian constraint.  Therefore, we expect this type of dynamics to occur in the vertex of the GFT.  Also, as we mentioned before, the states of the GFT have internal structure. So, the evolution of a state may change the combinatorics of the state for a given geometry and topology.  It is not implausible, consequently, that geometry information would occur in the interaction term of the GFT. From a more fundamental viewpoint, we recognize these GFTs as the static ultra-local limit of generalized GFTs \cite{Ori_gengft}.  This is a concept coming from standard QFT, where an underlying causal field theory gives rise to the projection onto the physical state space of an overlying theory, in some limit.

\section{The Ponzano-Regge model and the origins of conventional 3d group field theory} 
\label{3dprelim}

We now start the next step in our procedure.  We finished with the formal aspects of quantum gravity and we need to develop a rigorous definition of the $1st$ quantized theory of geometry.  This has existed for many years in a variety of guises.  We choose the discrete covariant version:  the spin foam model.  We shall only deal with the $3d$ model proposed by Ponzano and Regge for quantum gravity without cosmological constant \cite{Topological}.

\subsection{Classical Theory}
\label{3dclass}

To begin this section we recount, in brief, the pertinent features of classical $3d$ gravity.  We will confine ourselves to Riemannian signature and zero cosmological constant, for technical simplicity.  It is well known that the dynamics of this theory can be summed up by a $BF$-theory action.   
The basic variables in the classical theory are:  an $\su(2)$-valued triad frame field $E = E^{i}_{\mu} J_i dx^{\mu}$;  and an $\su(2)$-valued spin connection field, $A = A^{i}_{\mu} J_i dx^{\mu}$.\footnote{$J_i$ ($i = 0,\,1,\,2$) are the generators of an $\su(2)$ algebra satisfying $\tr(J_iJ_j)=2\delta_{ij}$ and $[J_i,J_j]=2i\gep_{ijk}J_k$.  Furthermore, the $\mu = 0,\,1,\,2$ are space-time indices.} 
From these fundamental fields we can construct other familiar quantities: the metric $g_{\mu\nu} = E^{i}_{\mu} E^{j}_{\mu} \gde_{ij}$, the curvature $F(A) = dA + A \wedge A$, the torsion $d_AE = dE + [A,E]$.
Most importantly , we can now write down the $1$st order action for $3d$ gravity which encodes the dynamical
information of the theory\footnote{$\gka = 4\pi G_N$ where $G_N$ is Newton's constant in $3d$. We shall choose units so that $\gka = 1$.}:
\be
\label{3daction}
 \cS_{\cM}[E, A] = \f{1}{4\gka} \int_\cM \tr(E \wedge F(A)),
 \ee
where $\cM$ is a three dimensional oriented smooth manifold, and $\tr$ denotes the Killing form on $\su(2)$.
For $\cM$ without a boundary, we get the classical equations of motion from varying $E$ and $A$
\be
\label{classeqn}
 E:\quad F(A) = 0,\quad\quad A:\quad d_AE = 0.
\ee
The first equation imposes flatness of the connection, and the second its compatibility with the triad field, also known as the torsion-free condition. Since in the quantum theory we will be interested in transition amplitudes we should deal also with the case of manifolds with boundaries. In this scenario, the second equation of motion only holds if the connection $A$ is fixed on the boundary $\pd\cM$.

Furthermore, at the continuum level we write down the symmetries of this action:
\bes
\label{contsymm}
 \textrm{Lorentz symmetry} 
 &\left\{ 
 \begin{array}{l} 
 A  \rar  k^{-1}dk + k^{-1}Ak\\
 E  \rar k^{-1}Ek
 \end{array}\right.
 & \quad\quad\textrm{parametrized by $k\in\SU(2)$}, \\
 \textrm{Translation symmetry}
 &\left\{
 \begin{array}{l}
 A \rar A \\
 E \rar E + d_{A}\phi
 \end{array}\right.\hphantom{xxxxx}\,
 & \quad\quad \textrm{parametrized by $\phi \in \su(2)$}.
\ees
The translation symmetry holds due to the Bianchi identity provided $\phi=0$ on the boundaries $\pd\cM$ (i.e. the translation symmetry is fixed there). As a final word on the subject the \lq Poincar\'e' symmetry is equivalent {\it on-shell} to diffeomorphism symmetry provided $\det(E)\neq0$.

\subsection{Kinematical State Space}
\label{kinematical}

To find the kinematical state space in the quantum regime, we start from a $2+1$ decomposition of the action. From there we can read off the properties of the classical phase space.  This is parametrized by $\bA$ and $\bE$, the pull-back of $E$ and $A$ respectively to a hypersurface $\gSi$.  The symplectic structure is defined by:
 \be
 \label{symp}
  \{\bA^i_a(x), \bE^b_j(y)\} = \gde^b_a \gde^i_j
  \gde^{(2)}(x,y).
 \ee
The first class constraints are  $d_{\bA}\bE=0\,$ and $F(\bA)=0$, known as the Gauss constraint and curvature constraint respectively.  The Gauss constraint generates infinitesimal $\SU(2)$ (\lq Lorentz\rq) gauge transformations while the curvature constraint generates translation symmetry. Unlike in $4d$ where one deals explicitly with a vector constraint and Hamiltonian constraint, in 3d these are lumped into one constraint.

 The kinematical Hilbert space $\cH_{kin}$ is defined to comprise of those states that lie in the kernel of the Gauss constraint above.  We choose a polarization of phase space so that the connection $\bA$ is the configuration variable, with $\bE$ its canonical momentum. Then, a basis for these states is given by $\SU(2)$ gauge-invariant cylindrical functions. These functions are known as spin-network states.  Such a state is built upon of a graph $\daleth\subset\gSi$.  The holonomy of the connection $\bar{A}$ along an edge $\ebs$ is assigned to that edge in a given representation of $\SU(2)$.  The gauge symmetry acts at the vertices $\vbs$ and so generates intertwiners among the representations incident at that vertex.  Explicitly it is given by
\be
\label{kinstate}
 \Psi_\daleth(\bA) =  \left\langle \bigotimes_{\ebs\in\daleth} D^{j_{\ebs}}(g_{\ebs}(\bA)) \Bigg| \bigotimes_{\vbs\in\daleth} C^{j_{\vbs_1}\dots j_{\vbs_n}} \right\rangle.
\ee  
This space of states can be completed to a Hilbert space by suitable use of the Haar measure. We should note that at a trivalent vertex the intertwiner is unique, but for higher valency we must specify an element in the basis of intertwiners\footnotemark.

\footnotetext{By specifying an intertwiner, we are in effect decomposing the higher valent vertex into a product of 3-valent intertwiners.}

\subsection{Imposing the curvature constraint}
\label{hamiltonian}

There are now two possible paths down which we could continue our investigation of $3d$ quantum gravity: canonical and covariant.  Since we are more interested in the covariant way, we shall concentrate on that approach.  The partition function for $3d$ gravity is given formally by:
\be
\label{partition}
 \cZ_{GR} = \int \cD A \cD E\,
  e^{i\cS_{\cM}[E, A]}
 = ``\int\cD A\, \gde(F(A))"
\ee
We can make this formula rigorous by regularizing the path integral.  In the Ponzano-Regge model one regularizes using a simplicial lattice and replaces the variables by discrete analogues.  The manifold $\cM$ is replaced by a simplicial counterpart $\dman$, which has tetrahedra $t$, faces $f$, edges $e$ and vertices $v$. The dual $2$-skeleton $\dmans$, is an important structure for our regularization.  It is defined as the set of vertices $\vs$ ($\sim t$), edges $\es$ ($\sim f$)  and faces $\fs$ ($\sim e$)\footnote{Here, $\vs \sim t$ means that $\vs$ is dual in the topological sense to t.}. In the case of a manifold with boundary $\pd\cM$, the intersection of $\dmans$ with the boundary gives the boundary triangulation , and likewise the intersection of $\dmans$ with the boundary gives the dual graph on the boundary. We denote these by $\pd\dman$ and $\pd\dmans$ respectively.  The continuous fields are replaced as follows:
 \bes
 \label{discvar}
   E &\rar& \int_{e} E = X_e \in \su(2),\\
   A &\rar& \int_{\es} A = g_{\es} \in \SU(2),\\
   F(A) &\rar& \prod_{\es\subset \pd \fs} g_{\es}^{\gep_{\fs}(\es)}
    = G_e \in \SU(2).
\ees
where $\gep_{\fs}(\es) = \pm 1$ is the the relative orientation of the edge $\es$ and the face $\fs$. Thus the partition function assumes the form:
 \be
 \label{PRpart}
 \begin{split}
  \cZ_{PR}
  &= \prod_{\es}\int_{\SU(2)} dg_{\es} \prod_e \int _{\su(2)} dX_e\,\, e^{i\sum_{e}\tr(X_eG_e)}\\
  &= \prod_{\es}\int_{\SU(2)} dg_{\es} \prod_e \gde(G_e) = \Big(\prod_e
  \sum_{j_e}\Big) \Big(\prod_e d_{j_e}\Big)\prod_t
  \left\{
  \begin{array}{ccc}
  j_1 & j_2 & j_3\\
  j_4 & j_5 & j_6
  \end{array}
  \right\}_t,
 \end{split}
 \ee
after Plancherel decomposition of the $\gde$-functions\footnotemark. In this way, we introduce the $\su(2)$ representations to each face $\fs$ and an intertwiner to each edge $\es\, (\sim f)$. For four faces $f$  forming a tetrahedron $t$, these intertwiners combine to form a $6j$-symbol.

\footnotetext{Actually, there is a subtlety here \cite{FreLou_PR1}, in that the integral over $\su(2)$ evaluates to,
\be
\int d^3X\, e^{i\tr(Xg)} = 4\pi (\gde(g) + \gde(-g)).\nonumber
\ee
To obtain the results in the main text we must include a factor of $\frac{1+\gep(g)}{2}$ where $\gep(g) = \textrm{sign}(\cos\,\gth)$ and $g = \cos\, \gth + i\vec{\gsi}\cdot\vec{n}\sin\,\gth$.  We do not include this factor for simplicity.}

To see the boundary states arising naturally from the spin foam, we consider its intersection with a boundary and the subsequent labeling of the discrete structures $\pd\dmans$ and $\pd\dman$. A boundary edge $\ebs$ inherits the representation of the incident bulk face $\fs$ while a boundary vertex $\vbs$ inherits the intertwiner from the incident bulk edge $\es$.  Thus, in the covariant formalism, the boundary states are given by spin-networks.  This is comforting as we want the spin foam to impose the curvature constraint.

\subsection{Covariant symmetries and Gauge-fixing}
\label{fixing}

The Ponzano-Regge amplitude is only a formal topological invariant. The amplitude is divergent due to the infinite sum over representation value. It may be regularized by introducing a cutoff.  This can be done rigorously by deforming the algebra from $\SU(2)$ to $\U_q(\SU(2))$ where $q$ is some root of unit and is the cut-off parameter. This gives rise to the well-known Turaev-Viro model \cite{Topological}, an exact topological state sum which has been related to $3d$ quantum gravity with cosmological constant.

There is a reason for the divergence in the amplitude which is more intrinsic to the theory from which it originated.  It appeals directly to our experience  regarding gauge theory in general, i.e. the presence of gauge symmetries results in divergences in the path integral.  We should only integrate over gauge equivalence classes to get a sensible amplitude; this is done by gauge-fixing.  The continuum action has two symmetries:  Lorentz and translation, and although replacing the continuum manifold with a simplicial one destroys much of these symmetries, there is a residual action of each on the discrete manifold \cite{FreLou_diffeo, FreLou_PR1}. 

To ensure a finite amplitude, we shall use up the gauge freedom to fix elements of the amplitude to desired values.  For a systematic implementation of this procedure, we utilize two structures,  a maximal tree $T$ of edges in $\dman$ and a maximal tree $\Ts$ of edges in $\dmans$.  An exhaustive explanation of the subsequent procedure is given in \cite{FreLiv_noncompact}, but we can sum up the end result by saying that the gauge symmetry is used up to set every representation $j_e$ attached to $e\in T$ to zero and every holonomy $g_{\es}$ attached to $\es\in\Ts$ to the identity.   Gauge-fixing in such a manner gives rise to a Fadeev-Popov determinant which turns out in this case to be equal to $1$. Furthermore, the gauge-fixed amplitude turns out to be independent of the maximal trees $T,\, \Ts$ chosen.  In the presence of boundaries the maximal tree $T^*$ extends to edges $\ebs$ of
the graph $\pd\dmans$, but the tree $T$ does not.  In fact $T$ can have at most one vertex on the boundary.  The reason is that there is no translation symmetry on the boundary ($\phi = 0$ on $\pd\cM$).

For the Riemannian case that we have been dealing with the redundant Lorentz integration has nothing to do with the divergences as the $\SU(2)$ group is compact and has a normalized Haar measure\footnote{Such is not the case in the Lorentzian scenario where the gauge group is $\SU(1,1)$ (or $\SO(2,1)$) which is non-compact.}.  All the divergence is locked up in the redundant integration over the $\su(2)$-algebra variables $X_e$, $e\in T$. At the level of the amplitude this amounts to inserting a gauge-fixing observable \cite{FreLou_PR1} into the partition function (\ref{PRpart})
\be
\label{transob}
 \cO_T(j_e) = \prod_{e\in T}\gde_{j_e,0}.  
 \ee
Now this observable seems to destroy the flatness condition for the holonomy associated to edges in the maximal tree.  This redundancy of a maximal tree of $\gde$-functions may be seen directly at the level of the spin-foam amplitude.  The discrete Bianchi identity takes the form
\be
\label{bianchi}
 \prod_{e:v\subset\partial e} g_{\vs(e)}^{-1}G_{e}^{\epsilon_v(e)} g_{\vs(e)} = 1,
\ee
with $\gep_v(e)=\pm 1$ records the orientation of the edge $e$ with respect to the vertex $v$, and $g_{\vs(e)}$ is a specific product of group elements $g_{\es}$.  To explain this relationship in words, consider a vertex $v \in \dman$ and all the edges $e$ incident at $v$.  The Bianchi identity states that there exists an ordering of the edges $e$ such that the product of their associated holonomies (up to conjugation) is the identity element.  Now consider the non-gauge-fixed amplitude (\ref{PRpart})\footnote{We refer the reader to Appendix \ref{app_bianchi}.}.This forces the curvature to be zero on all edges e.  But for a vertex with $n$ incident edges, once $n-1$ are forced to be the identity, the Bianchi identity assures us that the final edge has zero curvature. This means that we have a redundant $\gde$-function.  This argument extends to a maximal tree.

\subsection{Inclusion of matter}

A recent advance in the spin foam formalism has been the inclusion of point particles in the Ponzano-Regge model \cite{FreLou_PR1, FreLou_PR2, FreLiv_PR3}.  The coupling of matter fields is obtained as anticipated, by treating a full Feynman graph $\gga$ of a particle field theory (of arbitrary spin), with its hidden dependence on geometric variables, as a quantum gravity observable.  The coupling between geometric and matter degrees of freedom at each line of propagation of the graph is obtained by a discretization of the continuum action describing the coupling of gravity to relativistic point particles in 3d, with the line of the Feynman graph thus being interpreted as the trajectory of a relativistic particle in a 3d space-time, and by the subsequent integration over particle data.

One considers a particle graph $\gga$ embedded into the space-time manifold ${\cal M}$.  The dynamical information
associated to this configuration is the minimal coupling of classical relativistic point particles to gravity. The action is given explicitly by:
\be
\label{matter_action}
  \cS_{\cM,\gga} =
  \frac{1}{4\gka} \int_\cM \tr(E \wedge F(A))
  + \int_\gga \tr((E + d_Aq) \wedge u(mJ_0)u^{-1})
  + \int_{\gamma} \tr(A \wedge u(sJ_0)u^{-1}).
 \end{equation}
The action maintains the same symmetries as in pure gravity as long as $u \rightarrow k^{-1}u,\, q \rightarrow k^{-1}qk$ under Lorentz
transformations and $u \rightarrow u,\, q \rightarrow q + \phi$ under translations.  The equations of motion are:
 \begin{displaymath}
 \label{mattereom}
 \begin{array}{ccc}
  E: & F(A) = p\delta_{\gamma}\,\, & \quad \quad p = m\, uJ_0u^{-1},\phantom{xxxxxxxxxxx}\\
  A: & d_{A}E = j\gde_{\gga} & \quad\quad j = s\, uJ_0u^{-1} - m[uJ_0u^{-1},q],
 \end{array}
 \end{displaymath}
where $\gde_{\gga}$ is the $\gde$-function with support on the worldline. The first two equations of motion report that the curvature and torsion are zero except on the worldline of the particle, where they are proportional to the momentum and angular momentum respectively.  The path integral is then
\begin{equation}
\label{matterpart}
  \cZ_{\cM,\, \gga} = \int \cD A \, \cD E\, \cD q\, \cD p\,\, e^{i\cS_{\cM,\, \gga}}.
 \end{equation}
Once again we provide a simplicial discretization of the manifold $\cM$ with certain edges adapted to the Feynman graph $\gga$, denoted by $\dman$ and $\gGa$ respectively.   Upon this structure, we replace the continuum geometric data as before with its discrete counterpart.  Here, we shall also do the same for the matter degrees of freedom which propagate along the worldline.  

Remember, we had five components to describe the particle: mass, spin, vector in the Cartan subalgebra, momentum and position (or alternatively: total angular momentum).  To this end, every edge $e\in\gGa$ in the particle graph is labeled by a deficit angle $m_e$, and a spin $s_e$.   The vector in the Cartan subalgebra is contain in a group element $h_{m_e}$\footnote{$h_{m_e} = e^{ m_e J_0}$ is an element of the $\U(1)$ Cartan subgroup of $\SU(2)$}.  Furthermore, the momentum of the particle is summed up in a variable $u_e\in\SU(2)$ colouring each edge\footnote{$u_e=S(\vec{u}_e)$ where $S$ is a
section $S:\SU(2)/\U(1)\rightarrow\SU(2)$. $\vec{u}_e$ is a vector on the unit 2-sphere.}. To the endpoints of each edge $e\in\gGa$,  we assign total angular momentum variables, $I_{s(e)}$, $I_{t(e)}$, where $s(e)$ and $t(e)$ are the source and target vertices respectively. These total angular momenta are representations of $\SU(2)$.

Thus-far, we have labeled the triangulation and its dual with the fundamental constituents. We construct the quantum amplitude out of these quantities.  For an edge $e\notin\gGa$, we impose flatness of the holonomy: $\gde(G_e)$. This is the usual case for pure gravity.  For an edge $e\in\gGa$, we force the curvature to be in the conjugacy class of $\gth_e$: $\delta(G_e\,u_e\, h_{m_e}\,u_e^{-1})$, where $h_{m_e}\in\U(1)$. This imposes the expected curvature deformation coming from the particle.  For an edge $e\in\gGa$, we also attach a spin projector:
\be
\label{spinproj}
 \gDe^{I_{t(e)}I_{s(e)}}_{s_e}(u_e)_{l_{t(e)}l_{s(e)}}
 =D^{I_{t(e)}}_{l_{t(e)}s_e}(u_e)D^{I_{s(e)}}_{s_el_{s(e)}}(u_e)^{-1}.
\ee
To every vertex $v\in\gGa$, we associate an invariant tensor $C^{I_{s(e)}\dots}_{l_{s(e)}\dots}$ intertwining the total angular momentum variables coming from each edge $e$ incident there.  For the group $\SU(2)$ these intertwiners are the Clebsch-Gordan coefficients.

First we summarize the kinematical properties of the theory.  A boundary state is defined on an open spin-network graph.  This has edges joining trivalent vertices, and edges joining a 4-valent vertex to the point where the particle punctures the boundary.  To the first type of edge we assign the holonomy of the connection in a given representation of $\SU(2)$.  To the second type of edge we allocate the holonomy of the connection in a given representation of $\SU(2)$ projected down onto the spin-$s$ component: $D^{j_{\ebs}}_{.s}(x_{\ebs})$, . We label the vertices with intertwiners.

The amplitude is well defined once we chose a gauge-fixing. In order to do so we choose $T$  a maximal tree of $\gDe/\gGa$ and $T^*$ a maximal tree of $\gDe^*$ and fix there as prescribed in \cite{FreLou_PR1}. Thus the particle amplitude may be written:
\be
\label{cpramp}	
 \cZ_{\dman,\,\gGa}=
 \int \prod_{e^*\notin T^* } dg_{e^*}
 \prod_{e\notin T \cup \gGa} \gde(G_{e}) 
 \prod_{e\in\gGa} \gDe( m_e) \int du_e\,
 \gde(G_{e}u_eh_{m_e}u_e^{-1})
 D^{I_{t(e)}}_{\cdot\,l_{t(e)}}(a_e)
 \gDe^{I_{t(e)}I_{s(e)}}_{s_e}(u_e)_{l_{t(e)}l_{s(e)}}
 \prod_{v\in\gGa}C^{I_{s(e)}\dots}_{l_{s(e)}\dots},
\ee
where $\gDe(m_e) = \sin m_e$ and $D(a_e)$ is a function of holonomies multiplying the spin projector.  Its origin is in the occurrence of the Bianchi identity when searching for momentum conservation at the vertices of $\gGa$. Remember that the Bianchi holds, up to conjugation of the holonomies.  Therefore, momentum is conserved up to conjugation and we need these elements in the spin projector to have a consistently defined amplitude.

We can reformulate this amplitude so as to make it more amenable to a field theory description.  There are two ways of thinking about this procedure: either as summing over gravity degrees of freedom so as to end up with an effective amplitude for the particle degrees of freedom; or as using the the topological nature of the state sum for pure quantum gravity in 3d to re-express the amplitude on the simplest possible discretization encoding the particles' degrees of freedom.  The simplest such diagram is such that the edges of $\gGa$ only, are the edges of the discretization.  In any case, we need the pure gravity sector to be topological to do this simply\footnotemark. 
\footnotetext{This will become important when we come to deal with the causal amplitudes.}
The amplitude becomes\footnote{
The following relations hold
\bes
 G&= &\cos\,\gth + i \vec{J}\cdot \vec{n}\sin\,\gth = \sqrt{1-  |\vec{P}|^2} + i \vec{\gsi}\cdot\vec{P},\\
  \int dG & = & \frac{1}{\pi^2}\int_{B_1} \frac{d^3\vec{P}}{\sqrt{1 - |\vec{P}|^2}},\quad\quad\quad \textrm{where $B_1$ is the unit ball,}\\
  \int du\,\, \gde(Guh_m u^{-1}) &=&\frac{\pi}{2} \frac{\cos\, m}{\sin m} \,\gde\left( |\vec{P}|^2 - \sin^2\,m\right),\\
  \gDe( m) & = & \sin\, m.
\ees
}, for spherical graphs
\be
\label{simpleamp}
 \cZ_{\gGa}=
 \int\prod_e dG_e\,
 \gde\left(|\vec{P}_e|^2-\sin^2 m_e\right)\,
 \gDe^{I_{t(e)}I_{s(e)}}_{s_e}\left(S\left(\frac{\vec{P}_e}{\sin m_e}\right)\right)_{l_{t(e)}l_{s(e)}}
 \prod_{v\in\gGa} \gde\left(\prod_{e: v\in\pd e} G_e^{\gep_v(e)}\right)
 C^{I_{s(e)}\dots}_{l_{s(e)}\dots},
\ee
where $\gep_v(e)$ registers the relative orientation of the incident edges at a vertex of the particle graph.  But we note that at the moment, we only have the Hadamard propagator for the coupled theory, and so we do not expect to be able to describe this in a conventional field theory setting.  It is, however, possible to derive this amplitude from a field theory.  We propose a field
\be
\tilde{\psi}_s\, : \, \SU(2) \rar \C\,\, ; \,\,\tilde{\psi}_s(g)
\ee 
with an action
\be
\label{matter_hadamard}
\begin{split}
\cS(\tilde{\psi}_s,\tilde{\psi}_s^*) =& \int dg\,\tilde{\psi}_s(g)\tilde{\psi}_s^*(g) + \frac{\gla}{3!} \int dg_1\,dg_2\,dg_3\, \tilde{\psi}_s(g_1)\,\tilde{\psi}_s(g_2)\,\tilde{\psi}_s^*(g_3)\, \gde\left(|\vec{P}(g_1)|^2-\sin^2 m\right)\, \gde\left(|\vec{P}(g_2)|^2-\sin^2 m\right)\\
 &\phantom{xxxxxxxx}\times\gde(g_3g_2g_1)\,
D^{I_1}_{n_1s}\left(S\left(\frac{\vec{P}(g_1)}{\sin\,m}\right)\right)\, 
D^{I_2}_{n_2s}\left(S\left(\frac{\vec{P}(g_3)}{\sin\,m}\right)\right)\, 
D^{I_3}_{n_3s}\left(S\left(\frac{\vec{P}(g_3)}{\sin\,m}\right)\right)
C^{I_1\,I_2\,I_3}_{n_1\,n_2\,n_3} + \textrm{c.c.},
\end{split}
\ee
where we have trivial dynamics in the kinetic term.  The Hadamard propagators are placed in the vertex term for the $\tilde{\psi}$ fields, but not the $\tilde{\psi}^*$ fields.  We only need it for one of the fields since the kinetic term ensures that a $\tilde{\psi}$ is always connected to a $\tilde{\psi}^*$.  The spin projectors are split in half as the particle is always present at two vertices.  We have restricted to trivalent vertices and one species of particle.   This does not look like any action that we would use in standard field theory, but we will see that it is very familiar from a group field theory setting.  Indeed, we expect that it arises as the effective field theory for matter on reducing the action occurring in \cite{OriRya_gftspinning}.

\subsection{Causality}
\label{causality}
In a recent article \cite{OriTla_causal}, Oriti and Tlas introduced the causal propagator for the Ponzano-Regge model.  For gravity in the first order formalism, causality entails a restriction on the triad field. We wish only positively oriented triads to contribute to the partition function etc.  Allowing the determinant of the triad to vary only in the non-negative sector assures such a restriction.  Once we pass to the discrete path integral, however, we must find a way to impose this constraint.  Essentially, the argument proceeds along the following lines.  The Ponzano-Regge amplitude takes the form (\ref{PRpart}):
\begin{equation*}
  {\cal Z}_{\dman} = \int \prod_{e^*\in\dmans}dg_{e^*} \prod_{e\in\dman} dX_e\,
  e^{i\sum_{e}\tr(X_eG_e)}.
\end{equation*}
Now writing out the variables explicitly in terms of their $\su(2)$ Lie algebra components: $X_e = \vec{x}_e \cdot \vec{J}$, $G_e = \cos\theta_e 1 + i\vec{n}_e\cdot \vec{J} \sin\theta_e$.  Thus we may rewrite the amplitude as:
\be
\label{PRangle}
 \cZ_{\dman} = \int \prod_{\es} dg_{\es} \prod_e dX_e\,
 e^{i\sum_{e} \vec{x}_e \cdot \vec{n}_e \sin\,\gth_e}
\ee
$\vec{x}_e \cdot \vec{n}_e$ is the proper-time in this case and thus integrating over all values of $\vec{x}_e \cdot
\vec{n}_e$ gives rise to the Hadamard function.  But integrating over $\vec{x}_e \cdot \vec{n}_e \geq 0$, we end up with the causal propagator. The resulting propagator takes the closed form 
\be
\label{causal_pure}
\frac{i}{(\sin\,\theta_e + i\epsilon)^3}
\ee
where $i\epsilon$ is the usual regularizing factor which occurs in a Feynman propagator\footnote{There is an important issue pertaining to the computation of this integral.  The integral needs to be gauge-fixed.  This has not been done in \cite{OriTla_causal}, but we refer the reader there for more details.}. Thus, the causal Ponzano-Regge amplitude takes the form:
\be
\label{CPRpart}
 \cZ_{\dman}^{C} = \prod_{e^*}dg_{e^*} \prod_{e} \frac{i}{(\sin\, \theta_e + i\epsilon)^3}
\ee
The amplitude is not a triangulation invariant, and gauge-fixing yields a Fadeev-Popov determinant which is not equal to 1.  This is anticipated, since the causal amplitude is not a solution to the Hamiltonian constraint.  It is still possible to fix the Lorentz symmetry, which has Fadeev-Popov equal to 1, as in the a-causal case.

We may also extend this to the coupled matter amplitudes.  Limiting the range of integration as before we acquire the causal coupled propagator. We divide our work into two areas:  

\medskip

\paragraph{Zero total angular momentum}

For the case of the spinless particle, we perform a similar procedure to obtain the following expression for the propagator
\be
\label{causal_matter}
\frac{\cos\,\gth_e}{\sin\,  m_e}\,\,\frac{i}{\sin^2\,\gth_e - \sin^2 \, m_e + i\gep_e}.
\ee
There is an important factor of $\cos\,\gth$ here which comes from treating causality immediately at the level of the spin foam partition function, and not waiting to impose causality at the level of the quantum amplitude (\ref{simpleamp}) by analogy with standard matter field theory. 
This replaces the Hadamard function occurring in (\ref{cpramp}) to give us 
\be
\label{causal_amp}
\cZ_{\dman,\,\gGa}^{C}=  \int \prod_{e^* \in \Delta^*}\int   d g_{e^*}  \prod_{e \in\dman/ \Gamma} \frac{i}{(\sin\,\theta_e + i\epsilon)^3}  \prod_{e \in \Gamma} \cos\,\theta_e \, \frac{i}{\sin^2\,\theta_e -\sin^2\, m_e + i \gep_e}
\ee
To get down to an amplitude which may be generated from a field theory, we need to re-insert the a-causal amplitude for the pure gravity sector.   This is because we need the triangulation invariance and the gauge-fix-ability of the original $\gde$-function amplitude in order to simplify to 
\be
\label{causal_simpleamp}
\cZ_{\gGa}^{C} =  \int \prod_{e\in\gGa} dG_e\, \frac{i\,\cos\,\gth_e}{|\vec{P}_e|^2 - \sin^2\,m_e + i \epsilon_e} \prod_{v\in\gGa} \gde\left(\prod_{e: v\in\pd e}G_e^{\gep_v(e)}\right).
\ee
We think that this is a valid approximation since by far the dominant contribution to the partition function from the pure gravity edges, comes from the configurations which are flat on those edges. This means that, as far as the particle edges are concerned, the only contributions that are relevant are those for which the connection is flat everywhere except at the particle graph. We see then that (\ref{causal_simpleamp}) arises as the Feynman amplitude of a group field theory 
\be
\label{causal_eft}
\cS(\tilde{\phi}) = \frac{1}{2} \int dg\,\,\frac{  |\vec{P}(g)|^2 - \sin^2\, m}{\sqrt{1 -  | \vec{P} |^2}} \,\tilde{\phi}(g)\, \tilde{\phi}(g^{-1}) + \frac{\lambda}{3!} \int dg_1\, dg_2\, dg_3\, \tilde{\phi}(g_1)\, \tilde{\phi}(g_2)\, \tilde{\phi}(g_3)\, \delta(g_3 g_2 g_1) .
\ee
where we restricted to trivalent particle graphs.  This field theory, in contrast with (\ref{matter_hadamard}), looks in overall form, like a standard field theory.  Indeed, in the low gravity limit, it tends to the Klein-Gordon action.

\medskip

\paragraph{Non-zero total angular momentum}

In the general instance of spinning particles, we need first to make a distinctive clarification.  In the classical theory and the on-shell quantum theory, there is no difference between momentum $uh_mu^{-1}$ we assign the particle from the start, and the momentum $F$ which comes from the gravity sector.  But this is not true in the causal case.  Here, the former momentum is still always on-shell, where as the latter is free to vary off-shell \cite{OriTla_causal}.  Therefore, the momentum coming from the gravity sector is the true quantum mechanical momentum of the particle.  The momentum $uh_mu^{-1}$ may then be thought of as an auxiliary momentum.  It is, therefore, important to make this substitution in the path integral before the integral over $u$ is made.   Hence, we propose that the propagator occurring in the spin foam amplitude is 
\be
\label{causal_spinning}
\cP^{s}(G_e) = \frac{\cos\,\gth_e}{\sin\, m_e}\,\,\frac{i}{\sin^2\,\gth_e - \sin^2 \, m_e + i\gep_e} \,\,
\gDe^{I_{t(e)}I_{s(e)}}_{s}\left(S\left(\frac{ |\vec{P_e}|}{\sin\,m_e}\right)\right)_{l_{t(e)}l_{s(e)}}.
\ee
We then perform the same procedure as for the amplitudes with zero total angular momentum.  Now the effective field theory to generate these amplitudes is 
\be
\label{causal_spinningeft}
\begin{split}
\cS(\tilde{\phi}) =& \frac{1}{2} \int \frac{dg }{\sqrt{1 -  | \vec{P} |^2}}\,\, \tilde{\phi}^{s}_{l_1}(g) \cK^s_{l_1l_2}(g)\tilde{\phi}^{s}_{l_2}(g^{-1})\\ 
&\phantom{xx}
+ \frac{\lambda}{3!} \int dg_1\, dg_2\, dg_3\,\, \delta(g_1 g_2 g_3) \,\,
(\vec{P}^{(k_1}(g_1)\tilde{\phi}^{s)})_{l_1}(g_1)\times \dots \times (\vec{P}^{(k_3}(g_3)\tilde{\phi}^{s)})_{l_3}(g_3) C^{I_1\,I_2\, I_3}_{l_1\, l_2\, l_3}.
\end{split}
\ee
where $(\cK^{s})^{-1} = \cP^{s}$. This is the same as the effective field theory occurring in \cite{KarLivOriRya_spinning}. $\vec{P}^{(k}(g)\tilde{\phi}^{s)}$ is the product of $k$ copies of the momentum related to $g$ and the spin-$s$ field fully symmetrized.  $k$ refers to the orbital angular momentum of the particle and its total angular momentum is $I = k+s$.

As a final point, we are left with the lingering question, as to why we should treat the gravity variables causally when we can just separate them off in such a manner and not do so.  We shall answer this in the section \ref{new}.

\subsection{Group field theory to date}
\label{convgft}

Group field theories (GFTs) arise as a generating functionals for spin foams. They emerged from considerations in $4$-dimensional simplicial models of quantum space-time.  In that case, the classical theory is not topological and proposals to quantize the theory on a finite simplicial lattice space-time, not only truncate the infinite set of gravitational degrees of freedom down to a finite set, but also break the background independence of the theory.  By summing over all lattice space-times, we can accommodate  an infinite set of degrees of freedom and reintroduce the property of background independence.  Indeed, one may go further to say that we do not even have a dependence on space-time topology.

Here, we recount how this sum can be realized as the sum over Feynman diagrams of a quantum field theory living on a suitable group manifold, with each Feynman diagram defining a particular lattice space-time. From one aspect, this is an extension of the matrix models of 2d dimensional quantum gravity with dynamical topology, or Òzero dimensional string theoryÓ \cite{Matrix}.  The field theory in question is that one proposed by Boulatov \cite{Bou_gft}. Boulatov showed that the Feynman expansion of a certain field theory over three copies of $\SU(2)$ generates triangulations, colorings and  amplitudes of the Ponzano-Regge formulation of 3d quantum gravity (and, taking the q deformation of SU(2), the amplitudes of the Turaev-Viro model).

We shall talk about some group field theories that occur in the literature, for  3d gravity both with and without matter coupling.  The first group field theory for 3d gravity was proposed by Boulatov \cite{Bou_gft} It has a field defined over three copies of $\SU(2)$
\be
\label{field}
\phi\, :\,  \SU(2) \times \SU(2) \times \SU(2) \rar \,\C \,\,; \,\, \phi(g_1,g_2,g_3)
\ee
The field is invariant under two symmetries.  The first is classed as a permutation symmetry: an invariance under even permutations of the arguments.  We impose this by projection
\be
\label{gftperm}
P_{\gsi}\field{1}{2}{3}  =  \sum_{\gsi\in S_3} \field{\gsi(1)}{\gsi(2)}{\gsi(3)}.
\ee
It assures that we generate a sum over all oriented manifolds.  The second is Lorentz symmetry, which is imposed by projecting onto the $\SU(2)$ invariant part of the field.
\be
\label{gftlor}
P_{g}\field{1}{2}{3}  =  \int_{\SU(2)} dg\, \phi(g_1\, g,  g_2\, g,  g_3\,g).
\ee
We might refer to this as a right shift symmetry elsewhere in the text.

\begin{figure}[h]
\centering
\psfrag{a}{$g_1$}
\psfrag{b}{$g_2$}
\psfrag{c}{$g_3$}
\psfrag{d}{$j_1$}
\psfrag{e}{$j_2$}
\psfrag{f}{$j_3$}
\includegraphics[width = 7cm]{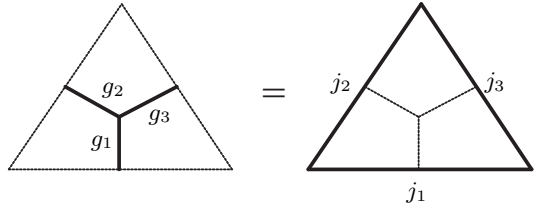}
\caption{Graphical description of the field $\phi$}
\label{oldfield}
\end{figure}

The field can be represented graphically by a triangle.  The $g_i$ arguments are the holonomies along edges of the spin network dual to this triangle.  As mentioned in the introduction this is the momentum representation.  We can Fourier transform to configuration representation, where the variables are representation labels.   Graphically, we associate the representations to the edge-length of the triangle and Lorentz invariance ensures that the edge-lengths satisfy the triangle inequalities Figure \ref{oldfield}

\begin{figure}[h]
\centering
\psfrag{a}{$g_1$}
\psfrag{b}{$\bar{g}_1$}
\psfrag{c}{$g_2$}
\psfrag{d}{$\bar{g}_2$}
\psfrag{e}{$g_3$}
\psfrag{f}{$\bar{g}_3$}
\includegraphics[width = 6cm]{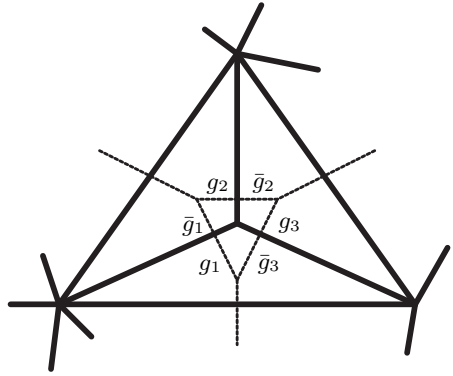}
\caption{Portion of boundary state}
\label{oldboundary}
\end{figure}

\begin{figure}[htbp]
\centering
\psfrag{A}{$\frac{\gla}{3!}$}
\psfrag{BBB}{$=0$}
\includegraphics[width = 7cm]{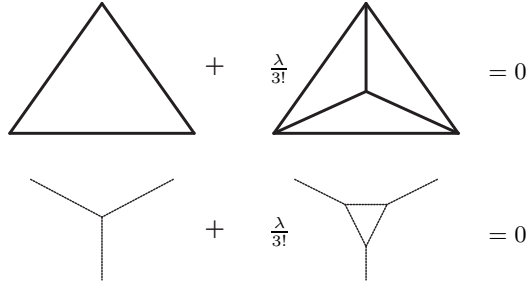}
\caption{Classical equation of motion}
\label{classical}
\end{figure}

The boundary observables for the GFT are constructed by the projecting the boundary information coming from the spin foam model onto a collection of GFT fields.  We illustrate our method on a portion of a boundary graph Figure \ref{oldboundary}. 
The associated spin-network label is 
\be
\label{netlabel}
C^{j_1j_3j_\cdot}_{m_1n_3 k_\cdot}\, C^{j_3j_2j_\cdot}_{m_3n_2 k_\cdot}\, C^{j_2j_1j_\cdot}_{m_2n_1 k_\cdot}\, D^{j_1}_{m_1n_1}(g_1\bar{g}_1^{-1})\, D^{j_2}_{m_2n_2}(g_2\bar{g}_2^{-1})\, D^{j_3}_{m_3n_3}(g_3\bar{g}_3^{-1}).
\ee
By projecting onto a collection of GFT fields we mean the following
\be
\label{gftobs}
\begin{split}
&\int dg_1\, dg_2\, dg_3\, d\bar{g}_1\, d\bar{g}_2\, d\bar{g}_3\, \phi(g_1, \bar{g}_3, g_\cdot)\, \phi(g_2, \bar{g}_1, g_\cdot)\, \phi(g_3, \bar{g}_2, g_\cdot)\\
&\phantom{xxxxxxxxxxxxxxxxxxxxxxxxxx}\times\, C^{j_1j_3j_\cdot}_{m_1n_3 k_\cdot}\, C^{j_3j_2j_\cdot}_{m_3n_2 k_\cdot}\, C^{j_2j_1j_\cdot}_{m_2n_1 k_\cdot}\, D^{j_1}_{m_1n_1}(g_1\bar{g}_1^{-1})\, D^{j_2}_{m_2n_2}(g_2\bar{g}_2^{-1})\, D^{j_3}_{m_3n_3}(g_3\bar{g}_3^{-1}).
\end{split}
\ee
The dynamics of this GFT are summed up by its action
\be
\label{gftaction}
\cS(\phi) = \int \prod_{i} dg_i\, \phi(g_1,g_2,g_3) \phi(g_1,g_2,g_3) + \frac{\lambda}{4!}\int \prod_{i} dg_i\, P_{g_a}\phi(g_1,g_2,g_3) P_{g_b}\phi(g_3,g_5,g_4)\, P_{g_c}\phi(g_4,g_2,g_6)\, P_{g_d}\phi(g_6,g_5,g_1).
\ee
The classical dynamics of this GFT are invested in the Euler-Lagrange equations.  For this group field theory this says 
\be
\label{gftclassical}
\phi(g_1,g_2,g_3) + \frac{\lambda}{3!} P_{g_b}\phi(g_3,g_5,g_4)\, P_{g_c}\phi(g_4,g_2,g_6)\, P_{g_d}\phi(g_6,g_5,g_1) = 0.
\ee
In diagrammatic form, Figure \ref{classical}, we see that this evolves a spin-network by a $1-3$ move. 
This is exactly the development of a spin-network state by the Hamiltonian constraint in loop quantum gravity.  Thus the classical dynamics of group field theory incorporates the quantum dynamics of loop quantum gravity. We shall devote more attention to the quantum dynamics, however, since we are interested in developing the $2nd$ quantized formalism. 

\begin{figure}[h]
\centering
\psfrag{a}{$g_1$}
\psfrag{b}{$g_2$}
\psfrag{c}{$g_3$}
\psfrag{d}{$u$}
\psfrag{S}{$s$}
\includegraphics[width = 3cm]{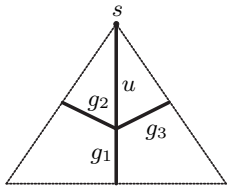}
\caption{Matter coupled to gravity field}
\label{coupledfield}
\end{figure}

The quantum dynamics are housed in the transition amplitudes.  Since we know the form of boundary observables, we shall concentrate on the partition function.  We shall restrict attention to the perturbative regime.  A given Feynman diagram of this theory is dual to a 3d simplicial manifold endowed with a flat geometry and the appropriate a-causal Ponzano-Regge quantum amplitude.  
\be
\label{gftpart}
\cZ^{GFT} = \sum_{\dman}\frac{\gla^{v[\dman]}}{\textrm{sym}[\dman]}\cZ_{\dman},
\ee
where $\cZ_{\dman}$ is the non-gauge-fixed Ponzano-Regge amplitude for the given simplicial complex $\dman$. $\vs[\dman]$ is the number of vertices in the Feynman graph and $\textrm{sym}[\dman]$ is the Feynman graph symmetry factor. We observe the validity of the previous statement by looking at the building blocks of a Feynman graph amplitude: the propagator and interaction operator
\bes
\label{gftoperator}
\cP(g_i,\bar{g}_j) &=& \gde(g_1\bar{g}_1^{-1})\,\gde(g_2\bar{g}_2^{-1})\, \gde(g_3\bar{g}_3^{-1}),\\
\cV(g_i) &=& \frac{\lambda}{4!} \int dg_a\, dg_b\, dg_c\, dg_d\, 
\gde(g_1\, g_a^{-1} g_d \, \bar{g}_1^{-1})\, 
\gde(g_2\, g_a^{-1} g_c \, \bar{g}_2^{-1})\, 
\gde(g_3\, g_a^{-1} g_b \, \bar{g}_3^{-1}) \\
&\phantom{=}&\phantom{xxxxxxxxxxxxxxxxxx}\times\,
\gde(g_4\, g_b^{-1} g_c \, \bar{g}_4^{-1})\, 
\gde(g_5\, g_b^{-1} g_d \, \bar{g}_5^{-1})\, 
\gde(g_6\, g_c^{-1} g_d \, \bar{g}_6^{-1})\nonumber
\ees
We notice that the vertex operator generates a tetrahedron and labels wedges of the plaquettes $\fs$ with a $\gde$-function enforcing flatness of the holonomy around each wedge. Thus, upon gluing the tetrahedra using the propagator, we glue the wedges and finish with the Ponzano-Regge amplitude.  We see that the dynamics of the GFT, the flatness constraints, are contained in the vertex term, while propagator has trivial dynamics.   Consequently, the action occurring in \cite{OriRya_gftspinning} for matter coupling is 
\be
\label{gftspin}
\begin{split}
S(\phi,\psi_s) = & S(\phi) + \frac{1}{2}\int\prod^{3}_{i=1} dg_i\, du\;
\psi_{s}(g_1,g_2,g_3;u)\,\psi_{s}(g_1,g_2,g_3;u)\\
&\phantom{xxxxx}+\mu_3\int \prod^{6}_{i=1}dg_i\, du_a\, du_b\, du_c\;
P_{g_a}\psi_{s}(g_1,g_2,g_3u_a^{-1}hu_a;u_a)\, P_{g_b}\psi_{s}(g_4u_b^{-1}h^{-1}u_b,g_3,g_5;u_b)\\
&\phantom{xxxxxxxxxxxxxxxxxx} P_{g_c}\psi_{s}(g_6,g_4,g_2u_c^{-1}hu_c;u_c)P_{g_d}\, \phi(g_6,g_5,g_1)
 \gde(u_a^{-1}hu_au_b^{-1}h^{-1}u_bu_c^{-1}hu_c)\\
 &\phantom{xxxxxxxxxxxxxxxxxxxxxxx}\sum_{\substack{I_a,I_b,I_c\\ n_a,n_b,n_c}}
 D^{I_a}_{sn_a}(u_a)\, D^{I_b}_{sn_b}(u_b)\, D^{I_c}_{sn_c}(u_c)C^{I_a\,I_b\,I_c}_{n_an_bn_c}
\end{split}
\ee
where we have only included the trivalent matter interaction.  We see lucidly from the action that the kinetic operator of this GFT has trivial dynamics once more.  Information about matter propagation and matter interaction is accommodated in the vertex. Before we look at this in more detail,  let us examine the new field.  It has four arguments, three are the usual gravity arguments, while the fourth contains information about the momentum of the particle.  Kinematically, the field represents a triangle on the boundary where the particle punctures it at one vertex.  The element $u$ represents the holonomy along the extra edge from the 4-valent vertex to the endpoint vertex Figure \ref{coupledfield}.
    
We see from the interaction term that momenta $u^{-1}hu$ are inserted at the in the gravity sector, and thus particles propagate along edges of the simplicial complex $\dman$.  Furthermore, we have inserted half of three spin-projectors and an intertwiner for the total angular momenta.  The other half of the spin projector comes from the adjacent tetrahedron.  What is important to note is that this field theory does not look like the effective field theory for matter in Section \ref{causality}, but it does have the same form as that of the matter field theory with Hadamard propagators in the vertex term (\ref{matter_hadamard}).  We expect that these field  is related in some more exact fashion, but we do not provide that connection here.  But, we claim that the GFT to which we coupled matter so far is not suited to achieving the causal effective field theory for matter in some sector.  Therefore, to remain this formalism we must move to the generalized GFT setting to impose causality. 

We, on the other hand, seek another way to implement causality in the group field theory.  In order for the group field theory to produce the Feynman propagator from its a kinetic term, we need to be able to encode the holonomy around a full plaquette (not just a wedge) in the kinetic term.  This is not possible in the current formalism.   Therefore, we propose a new one.

\section{The new model}
\label{new}

\subsection{Shift in perspective}
\label{shift}

We now investigate our new model which stems from a shift in perspective from the conventional one.  We have nothing to say more about the classical theory, and the kinematical states still have a spin-network basis.  Our alteration begins with the discretization procedure, when covariantly quantizing.  There are several ways of apprehending the meaning of what we propose and we will list them here. Each has its benefits and since they are equivalent we can choose whichever one suits our needs.  

The first is to say that instead of discretizing the manifold using a triangulation, we use the more general structure of a $CW$-complex.  As we will be interested in generating these from a group field theory context we do not use the most general $CW$-complex, but only those whose topological dual is a simplicial complex. From this point of view, we can use the results of \cite{GirOecPer_topological}, which tell us that $BF$-theory based on these structures is equally valid as a topological invariant (when a suitable deformation of the algebra is used). 

Alternatively, one can perceive this method as replacing the continuum manifold by a triangulation, as per usual but instead we swap the importance of the triangulation and its dual. What we propose is that instead of discretizing the connection $A$ onto the edges $\es$, and the triad onto the edges $e$, we swap the structures.  Thus
\bes
\label{newdisc}
 e &\rar& \int_{\es} e = X_{\es},\\
 A &\rar& \int_e A = g_e,\\
 F(A) &\rar& G_{\es} = \prod_{e}g_e
\ees
Thus, we retain the \lq unstarred' labels as pertaining to the simplicial complex even though it is now the spin foam structure.  Furthermore, the \lq starred' labels still mark components of the $CW$-complex (also referred to as the cell complex) even though this is the discretizing structure. This is a more natural outlook when treating the group field theory, as we will still use tetrahedra as the fundamental building blocks. The discrete analogue of the partition function is
\be
\label{dualpart}
 \cZ_{\dmans} = \int \prod_e dg_e \prod_{\es} dX_{\es} \,
  e^{i\sum_{\es}\tr(X_{\es}G_{\es})}\\
  = \int \prod_edg_e\, \gde(G_{\es})
\ee
Note that unlike the analogous expression in (\ref{PRpart}) we do not explicitly decompose in terms of $\gde$-functions since the more general building blocks of the $CW$-complex are not necessarily tetrahedral and we would not get $6j$-symbols.  Of course, such a decomposition is still possible, and we note for future reference that such an operation leads to faces $f$ labeled by representations $j$ of $\SU(2)$ and edges $e$ coloured by intertwiners.  But, it is important to remember that we are not dealing with the Ponzano-Regge model anymore.   
So far we have the partition function, which sums up the vacuum-to-vacuum dynamics of the theory, but we need to define states and transition amplitudes in order to have a coherent quantum theory.  These follow the same transmogrification as the spin foam amplitudes.  The intersection of $\dman$ with the boundary $\pd\cM$ gives us the boundary triangulation $\pd\dman$, while a likewise intersection of $\dmans$ gives us $\pd\dmans$.  Thus, the edges $\eb$ inherit representations labels, and the vertices $\vb$ inherit the intertwiners of the incident faces $f$ and edges $e$ respectively.  The vertices now are not necessarily trivalent, but as we mentioned earlier, for higher valence vertices, we can chose an intertwiner from the basis.  Furthermore, we also get bivalent vertices, but there is also an intertwiner for these, the Kronecker delta.   We ascertain that spin-networks form the basis for the boundary states. 
\be
\label{new_states}
 \Psi_{\pd\dman}(\bA) =  \left\langle \otimes_{e\in\pd\dman} D^{j_e}(g_e(\bA)) | \otimes_{v\in\pd\dman} C^{j_{v_1}\dots j_{v_n}} \right\rangle.
\ee
Finally, we have the states, so it is straightforward to construct the transition amplitude, which is just
\be
\label{new_transamp1}
\langle \Psi_{\pd\dman_2} |  \Psi_{\pd\dman_1} \rangle = \int \prod_e dg_e \prod_{\es} dX_{\es} \,  \Psi_{\pd\dman_1}(\bA) \Psi_{\pd\dman_2}(\bA)
  e^{i\sum_{\es}\tr(X_{\es}G_{\es})}, 
\ee
where we pick the manifold and its subsequent discretization such that its boundary is compatible with the graphs upon which the states are based $\pd\dman_1,\, \pd\dman_2$.

Now the Lorentz symmetry acts at vertices $v$, and translation symmetry at the vertices $v^*$. And we must choose once again maximal trees in both $\dman$ and $\dmans$ upon which to fix this symmetry.

Matter can once again be coupled to gravity in the quantum regime.  As a matter of fact, the technique involved is no different from the previous one, except for one major difference.  The particles are thought of as propagating along edges of the CW-complex, thus  $\gGa \subset \dmans$.  Our amplitude therefore reads
\be
\label{new_matteramp}
\begin{split}
\cZ_{\dmans,\,\gGa}= &
 \int \prod_{e\notin T } dg_{e}
 \prod_{\es\notin \Ts \cup \gGa} \gde(G_{\es}) 
 \prod_{\es\in\gGa} \gDe( m_{\es}) \int du_{\es}\,
 \gde(G_{\es}u_{\es}h_{m_{\es}}u_{\es}^{-1})\\
 &\hphantom{xxxxxxxxxxxxxxxxxxxxxxxxxxxxxxxxxxx}\times
 D^{I_{t({\es})}}_{\cdot\,l_{t({\es})}}(a_{({\es})})
 \gDe^{I_{t({\es})}I_{s({\es})}}_{s_{\es}}(u_{\es})_{l_{t({\es})}l_{s({\es})}}
 \prod_{\vs\in\gGa}C^{I_{s({\es})}\dots}_{l_{s({\es})}\dots}.
\end{split}
\ee
Since the CW-complex has only 4-valent vertices, we have as a consequence that we can have at most a 4-valent interaction in the particle graph.  We will note later how this may be generalized.  Moreover, the causal amplitudes are a direct transfer also, just replacing the Hadamard functions by the Feynman propagators
\bes
 \cZ_{\dmans}^{C} &=& \prod_{e}dg_{e} \prod_{\es} \frac{i}{(\sin\, \theta_{\es} + i\epsilon)^3},\label{new_causal}\\
 \cZ^{C}_{\dmans,\,\gGa} &=&
 \int \prod_{e\notin T } dg_{e}
 \prod_{\es\notin \Ts \cup \gGa}\frac{i}{(\sin\, \theta_{\es} + i\epsilon)^3}
 \prod_{\es\in\gGa}\,
 \frac{i\cos\,\gth_{\es}}{\sin^2\,\gth_{\es} - \sin^2 \, m_{\es} + i\gep_{\es}} \nonumber\\
 &  &\hphantom{xxxxxxxxxxxxxxxxxxxxxxx}\times
 D^{I_{t({\es})}}_{\cdot\,l_{t({\es})}}(a_{\es})
 \gDe^{I_{t({\es})}I_{s({\es})}}_{s_{\es}}\left(S\left(\frac{|\vec{P_{\es}}|}{\sin\,m_{\es}}\right)\right)_{l_{t({\es})}l_{s({\es})}}
 \prod_{\vs\in\gGa}C^{I_{s({\es})}\dots}_{l_{s({\es})}\dots}. \label{new_causalmatter}
\ees
We now possess the essential ingredients at the level of the spin foams so we can progress onto the group field theory scenario.

\subsection{Group field theory}
\label{new_gft}

When we move to the field-theoretic side of any model theory, we can draw from considerable experience in standard quantum and statistical field theory.  There, the analysis is broken up into two parts, the kinematical side and the dynamical side. The kinematical side involves an examination of the field and its inherent symmetries, along with that of a set of fields defining a boundary state.  For the dynamics, we must scrutinize the action, which should reveal to us many more facets; the Euler-Lagrange equations allow us to see the classical solutions, we might hope that the rest of the symmetries would become clear but we have seen that this can be obscure.  Furthermore, its input to the path integral, should allow us to analyze the quantum dynamics, both non-perturbatively and perturbatively.  There are many interesting features that are worth testing: its Borel summability, its ability to allow for matter coupling, its semi-classical limit,  the investigation of the Feynman amplitudes.

\subsubsection{Kinematical}
\label{new_kinematical}
As with the field theories defined in Section \ref{convgft}, we wish to embark upon a systematic description of our new GFT.  In this section we shall develop its kinematical properties, i.e. the field and the boundary state.  Before we do this, however, let us recall, what we want from the spin foam arena.  At that level, the boundary states are defined upon graphs that are simplicial and furthermore, dual to a trivalent discretization of the boundary 2-manifold.

\medskip

\paragraph{Field and symmetries} It is natural, therefore, to take our field to be very similar in content to those of previously defined group field theories, but the information encoded by the field will rest on different structures.   We therefore define our field over three copies of the gauge group $\SU(2)$.
\begin{equation}
\label{new_field}
 \phi:\, \SU(2) \times \SU(2) \times \SU(2) \rightarrow
 \mathbb{C}\,;\quad \field{1}{2}{3},
\end{equation}
\begin{figure}[h]
\psfrag{a}{$g_1$}
\psfrag{b}{$g_2$}
\psfrag{c}{$g_3$}
\psfrag{efd}{$g_3g_2g_1$}
\centering
\includegraphics[width = 3cm]{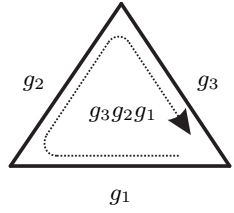}
\caption{New meaning of the $\phi$ field}
\label{newfield}
\end{figure}
The field represents a triangle in the spin network Figure \ref{newfield}, that is, three edges forming a plaquette, and the arguments $g_i$ are the holonomies along its edges.  The dual structure forms three edges and a vertex of the discretization of the boundary. The field does not possess a right shift symmetry as this is simply incompatible with the above interpretation of the field.  The Lorentz symmetry is now obscure at the level of the field but we will show a procedure later, on how to make it explicit.

Furthermore, translation symmetry is not observable at this stage either.  We expect this to be the case, since in our theory, translation symmetry is fixed on the boundary.  We are left with one final symmetry, which is reserved for group field theories and that is invariance under even permutations of the arguments.  This is required as usual, so that we may invoke the sum over all orientable manifolds in our Feynman diagram expansion.  We impose it by projection
\be
\label{new_perm}
P_{\gsi}\field{1}{2}{3}  =  \sum_{\gsi\in S_3} \field{\gsi(1)}{\gsi(2)}{\gsi(3)}.
\ee
Finally, complex conjugation of the field corresponds to an antisymmetric permutation
\be
\label{new_anti}
\phi^*(g_1,g_2,g_3) = \phi(g_3^{-1}, g_2^{-1}, g_1^{-1})
\ee
See \cite{DepPet_gft} for details.

\medskip

\begin{figure}[h]
\centering
\psfrag{a}{$g_1$}
\psfrag{b}{$g_2$}
\psfrag{c}{$g_3$}
\includegraphics[width = 5cm]{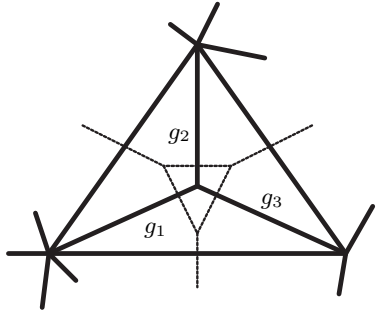}
\caption{Portion of boundary state}
\label{newstate}
\end{figure}

\paragraph{Boundary states} We see much more interesting structure appearing once we start to construct boundary states.  Such a state in our model is formed from a product of fields projected down onto the spin-network basis.  As a state may be arbitrarily complicated, we chose to illustrate our procedure on a portion of such a graph Figure \ref{newstate}. The spin network graph is simplicial and its label from the spin foam theory is
\be
\label{new_bound}
D^{j_1}_{m_1n_1}(g_1)\,  D^{j_2}_{m_2n_2}(g_2)\,  D^{j_3}_{m_3n_3}(g_3)\, C^{j_1\,j_2\,j_3}_{n_1\,n_2\,n_3},
\ee
where we have only labeled the central edges, the spokes.    
We now project the GFT fields associated to those triangles onto the spin network basis as follows
\be
\label{new_gftbound}
\int dg_1\, dg_2\, dg_3\, \phi(g_1^{-1},g_2,g_{\cdot})\, \phi(g_2^{-1},g_3,g_{\cdot})\, \phi(g_3^{-1},g_1,g_{\cdot}) \,D^{j_1}_{m_1n_1}(g_1)\,  D^{j_2}_{m_2n_2}(g_2)\,  D^{j_3}_{m_3n_3}(g_3)\, C^{j_1\,j_2\,j_3}_{n_1\,n_2\,n_3}
\ee
Notice that we have assigned an orientation to the faces and the edges so that $g_i$ appears in one field but $g_i^{-1}$ in the other.  Also Lorentz symmetry becomes manifest in the field and an $\SU(2)$ element $\gal$ is associated to the vertex at the centre.  Then, we may perform a transformation on the arguments of the fields
\be
\label{boundarylor}
\phi(g_i^{-1}, g_j, g_{\cdot}) \rar \phi(\gal^{-1}\,g_i^{-1},\,  g_j\,\gal,\, g_{\cdot}).
\ee
Then, redefining the arguments $g_i\, \gal \rar g_i$ leads to the element $\gal$ appearing in the representation functions which are intertwined by the $3j$-symbol. We can use the invariance of the intertwiner to demonstrate the invariance of the state.  Thus Lorentz invariance is locked into the field, it just requires more than one field to make it explicit.  

We have now finished the group field theory kinematics.  We defined the field and constructed the boundary observables.  We pass onto the dynamics.

\subsubsection{Dynamics}
\label{new_dynamics}

\paragraph{Action} Our first task in this section is to propose an action, as with any field theory.  Clearly our main motivation comes from the spin foam theory which we wish to characterize our Feynman graphs.  Thus, the basic building block in our perturbative expansion ought to be the tetrahedon.  In this instance, we employ a group field theory with the same basic structure as before.  But as with the previous section the constituents take on different meanings in our theory.  To clear our minds, the general structure of a GFT is
\be
\label{gen}
\begin{split}
\cS(\phi) =& \frac{1}{2} \int \prod_{i} dg_i\, d\bar{g}_i \phi(g_1,g_2,g_3)\,K(g_i, \bar{g}_j)\, \phi(\bar{g}_1,\bar{g}_2, \bar{g}_3)\\
 &\phantom{xxxxxxxxxxxx}+  \int \prod_i dg_i\, d\bar{g}_i\, V(g_i,\bar{g}_j)\, \phi(g_1, g_2, g_3)\, \phi(g_5,\bar{g}_3, g_4)\, \phi(g_6, \bar{g}_4, \bar{g}_2) \, \phi(\bar{g}_5, \bar{g}_6, \bar{g}_1).
\end{split}
\ee
Moreover, we want a generic Feynman amplitude (\ref{new_causal}) to be
\be
\label{new_transamp2}
 \cZ_{\dmans}^{C} = \prod_{e}\int dg_{e} \prod_{\es} \frac{i}{(\sin\,\theta_{\es} + i\epsilon)^3},
\ee
where $\cZ_{\dmans}$ is a product of amplitudes for the edges and vertices of the Feynman graph. This should help us decide of what the kinetic and vertex operators consist. Also, notice that we have chosen the causal version of the amplitude. Our reason will become apparent presently.  Looking at the amplitude, to each edge $e$ we assign a parallel transport and to each face $f \sim\es$ the causal version of flatness.  Now, the GFT field represents the holonomy around a face $f$. This leads us to define the propagator of our theory as the causal amplitude times some consistency conditions.  These conditions ensure the parallel transport variables assigned to coincident edges is the same.  Therefore, the vertex operator need not contain any interesting dynamical content, just similar consistency conditions.
\bes
\label{new_operators}
\cK(g_i, \bar{g}_j) &=& \frac{\sin^3\, \theta}{i}\,\,\gde(g_1\bar{g}_3)\,\gde(g_2\bar{g}_2)\,\gde(g_3\bar{g}_1), \quad\quad\quad \textrm{where} \quad\quad \gth = \gth(g_3g_2g_1),\\
\cV(g_i, \bar{g}_j) &=& \frac{\gla}{4!}\gde(g_1\bar{g}_1)  \gde(g_2\bar{g}_2)\,  \gde(g_3\bar{g}_3)\,  \gde(g_4\bar{g}_4)\,  \gde(g_5\bar{g}_5)\,  \gde(g_6\bar{g}_6).
\ees
This results in an action of the form
\be
\label{new_action}
\begin{split}
S(\phi) =& \frac{1}{2} \int \prod_{i} dg_i \, d\gal_i\, \phi(\gal_2^{-1}g_1\gal_1, \gal_3^{-1}g_2\gal_2, \gal_1^{-1}g_3\gal_3)\, \frac{\sin^3\, \theta}{i}\, \phi^*(\gal_2^{-1}g_1\gal_1, \gal_3^{-1}g_2\gal_2, \gal_1^{-1}g_3\gal_3)\, \\
 &\phantom{xxxxxxxx}+  \frac{\gla}{4!}\int \prod_i dg_i \prod_j d\gal_j\,  \phi(\gal_2^{-1}g_1\gal_1, \gal_3^{-1}g_2\gal_2, \gal_1^{-1}g_3\gal_3)\, 
\phi(\gal_1^{-1}g_5\gal_4, \gal_3^{-1}g_3^{-1}\gal_1, \gal_4^{-1}g_4\gal_3)\\
 &\phantom{xxxxxxxxxxxxxxx}\times
 \phi(\gal_{4}^{-1}g_6\gal_2^{-1}, \gal_3^{-1}g_4^{-1}\gal_4, \gal_2^{-1}g^{-1}_2\gal_3) \,
  \phi( \gal_4^{-1}g_5^{-1}\gal_1, \gal_2^{-1}g_6^{-1}\gal_4, \gal_1^{-1}g_1^{-1}\gal_2).
\end{split}
\ee
To explain the presence of the $\gal$ variables in the action we make clear some of category theory roots underlying the model. We refer the reader to \cite{Pfe_gauge} of the relevant definitions.  But it boils down to the fact that the model is based upon assigning a morphism $g$ to each edge of the simplicial complex and a natural transformation $\gal$, to each vertex.  The natural transformations act on the morphisms as in the action, and we ensure the symmetry by projection.  The natural transformations embody the Lorentz gauge invariance of the model.  The remnant of the translation symmetry is still obscure.  This is due to the fact that we simply substituted a causal propagator from the spin foam formalism.  The calculation of this propagator was not gauge fixed so we need to do this in order to rigorously define the kinetic operator.

\medskip

\paragraph{Quantum dynamics} In this section on the quantum dynamics we will deal with the perturbative expansion into Feynman diagrams so that one may eventually calculate transition amplitudes order-by-order.  The partition function takes the form
\be
\label{new_partition}
\cZ^{GFT} = \sum_{\dmans}\frac{\gla^{v[\dmans]}}{sym[\dmans]}\cZ^{GFT}_{\dmans}.
\ee
$\cZ^{GFT}_{\dmans}$ and $\cZ^{C}_{\dmans}$ coming from (\ref{new_causal}) are not identical as amplitudes.  There are infinities arising solely from the fact that we generated the amplitude directly from a GFT.  We can see immediately that these occur because the consistency conditions enforce every parallel transport attached to a given edge $e$ of the simplicial complex to be the same.  But in surveying all the faces $f$ sharing that edge we recognize that we enforce this consistency one time too many.  Thus we obtain an infinity for every edge $e$ of the simplicial complex (spin foam).  Saying this in another way, we have an infinite contribution to the amplitude for every face $\fs\sim e$, that is every loop of edges $\es$ in the cell complex (discretization).  We can explain how this occurs from a field theory point of view. It is very similar to the occurrence of momentum loops in standard field theory.  For a Feynman graph in standard quantum field theory there is an undetermined momentum attached to each loop.  Thus all one needs to do is to cut out enough edges so that there are no loops left.  That is we leave a maximal tree of edges in the Feynman graph. There are $|\vs| - 1$ edges in a maximal tree.  Thus, the number of infinities coming is $|\es| - |\vs| + 1$.  Now, for a Feynman graph in the group field theory, we have an undetermined position attached to each loop. And because of the intrinsic 3-dimensional setting of the cell complex we get an extra $|v| - 1$ infinities from the 3-cells in the cell complex.  This gives us a total of $|\es| - |\vs| + |v|$ which is equal to $|e|$ for a closed manifold.  Once again, in this scenario, these infinities do not come into play if we restrict to tree level graphs.  But we know from standard field theory that the classical solution to the full field theory with sources is given by the sum over tree level graphs. 

In tandem with $\cZ^{C}_{\dmans}$, we have not gauge-fixed the group field theory amplitudes, so they contain infinities from over-counting the gauge symmetries. (Remember that the translation symmetry is broken along the \lq time-like' direction by the causality restriction but the other two components still parametrize a non-compact symmetry.)

Finally, since the causal amplitudes are not invariant under Pachner moves, let alone the more general moves needed to relate CW-complexes, we cannot say much about how the sum over Feynman graphs of this GFT relates to the sum over Feynman graphs in the other causal GFTs.  But we can show that were we to use this formalism to generate e.g. Ponzano-Regge-like (more precisely Turaev-Viro) amplitudes, our sum over graphs would neither restrict nor enlarge the theory.  Such an amplitude defined on a $CW$-complex can always be rewritten, by a sequence of moves shown in \cite{GirOecPer_topological}, as an equivalent amplitude on a simplicial complex.  For the converse we must show that the amplitude defined on a simplicial complex can always be rewritten ona cell complex whose dual is a simplicial complex.  To do this, we consider such a simplicial complex, and its dual $CW$-complex. Forgetting about the amplitude for the moment, this $CW$-complex, may be refined to a simplicial complex, which is in turn dual to a $CW$-complex.  Now consider the amplitude defined on this latter CW complex.  By a sequence of moves we may relate this to the original simplicial complex and we are done.  Of course the coefficients multiplying each separate geometry and topology are different in the two summations.  More importantly,  the Ponzano-Regge amplitude does not distinguish among non-flat geometries so we have not shown that we sum over all geometries in our new GFT.  We leave this for future investigation.  

Our construction provides another motivation, complementing \cite{LivOri_causal, OriTla_causal, Tei_qmechanics, Ori_feynman, Ori_gengft}, to consider causality arising from orientation as the most basic form of quantum gravity, since if we subscribe  to the fundamental nature of the GFT approach, and that it should arise as a $2nd$ quantization in an analogous fashion to matter field theory, then we see that in its formulation we must include causality from the start, and the a-causal amplitudes arise from replacing the propagator in the Feynman expansion.

\paragraph{Free field theory} Now let us look at the free field sector of the GFT. Its action is 
\be
\label{new_free}
\cS_{free}(\phi) =  \frac{1}{2} \int \prod_{i} dg_i \, d\gal_i\, \phi(\gal_2^{-1}g_1\gal_1, \gal_3^{-1}g_2\gal_2, \gal_1^{-1}g_3\gal_3)\, \frac{\sin^3\, \theta}{i}\, \phi^*(\gal_2^{-1}g_1\gal_1, \gal_3^{-1}g_2\gal_2, \gal_1^{-1}g_3\gal_3)
\ee
which gives the classical equation of motion
\be
\frac{\sin^3\, \theta}{i}\, \phi^*(g_1, g_2, g_3) = 0
\ee
In analogy with matter field theory, a solution to the equation of motion is the corresponding Hadamard function: $\phi(g_1,g_2,g_3) \sim \gde(g_3g_2g_1)$, which is exactly the a-causal propagator occurring in previous models of quantum space-time.  Thus, we can explain our reason for choosing the causal amplitude over the a-causal amplitude.  This is rather obvious from the usual field theory mind set but not so much in the group field theory where a-causal amplitudes are the norm. It serves, however, to confirm that our model is akin to standard field theory because the a-causal propagator is the Hadamard function for our newly defined equations of motion; it is not a Green's function; it is not invertible as an operator on field space. It also suggest that flat geometries are the classical solutions to our group field theory, and thus exactly the answer we are looking for in loop quantum gravity.  

Finally, this acts like a free field theory should.  It propagates an in-state to an out-state without any alteration.  It is completely gauge-fixed as an amplitude and the Feynman graph it generates has the topology $\gSi\times\R$ where $\gSi$ is the manifold for which $\pd\dman$ is a discretization.  Thus, it is related to a subset of diagrams coming from the canonical loop quantum gravity approach. Now looking at the interacting theory, we see that it is responsible for the interaction of boundary states.  Now, since we have placed all of the causal dynamics into the kinetic term, the vertex term is directly responsible for spatial topology change, by allowing for the interaction of the local simplicial chunks of space.  It is also responsible for the construction of non-trivial space-time topologies.

\subsubsection{Matter Coupling}
\label{new_matter}
This provides a startlingly clear motivation for our GFT.  The effective field theory for matter outlined in Section \ref{causality} is an example of a very simple group field theory.  But it is a group field theory which implicitly includes all the gravity information that we need to describe point particles propagating and interacting on a quantum gravity background.  Thus, what we need to do is release quantum gravity information from this GFT.  But, its causal dynamical information lies in the propagator, so we really would not expect the group field theories of Section \ref{convgft} to be directly related. Our group field theory, however,  keeps its causal dynamics in the kinetic term.  Thus we have a chance.  

We look to the spin foam amplitudes as an initial guide.  The generic spinning case comes as (\ref{new_causalmatter})
\bes
\cZ^{C}_{\dmans,\,\gGa} &=&
 \int \prod_{e\notin T } dg_{e}
 \prod_{\es\notin \Ts \cup \gGa}\frac{i}{(\sin\, \theta_{\es} + i\epsilon)^3}
 \prod_{\es\in\gGa}\,
 \frac{i\cos\,\gth_{\es}}{\sin^2\,\gth_{\es} - \sin^2 \, m_{\es} + i\gep_{\es}} \nonumber\\
 &  &\hphantom{xxxxxxxxxxxxxxxxxxxxxxx}\times
 D^{I_{t({\es})}}_{\cdot\,l_{t({\es})}}(a_{\es})
 \gDe^{I_{t({\es})}I_{s({\es})}}_{s_{\es}}\left(S\left(\frac{|\vec{P_{\es}}|}{\sin\,m_{\es}}\right)\right)_{l_{t({\es})}l_{s({\es})}}
 \prod_{\vs\in\gGa}C^{I_{s({\es})}\dots}_{l_{s({\es})}\dots}. \nonumber
\ees
In the case of matter with zero total angular momentum this amplitude takes the simpler form 
\be
\label{new_causalspinless}
 \cZ^{C}_{\dmans,\,\gGa} =
 \int \prod_{e\notin T } dg_{e}
 \prod_{\es\notin \Ts \cup \gGa}\frac{i}{(\sin\, \theta_{\es} + i\epsilon)^3}
 \prod_{\es\in\gGa}
 \frac{i\cos\,\gth_{\es}}{\sin^2\,\gth_{\es} - \sin^2 \, m_{\es} + i\gep_{\es}}\\
 \ee

\medskip

\paragraph{Field and symmetries} We get going on this field theory by defining a new field to represent gravity coupled to a generic spin-$s$ particle. We specify the field over three copies of $\SU(2)$ 
\be
\label{new_cfield}
\psi^{s}\, :\, \underbrace{\SU(2)\times \dots\times\SU(2)}_{3}\, \rar\, \C\,\, : \,\, \psi^{s}_l(g_1,g_2,g_3).
\ee 
The group is itself valued in the spin-$s$ representation of $\SU(2)$.  The three arguments have the same interpretation as gravity parallel transports as before.  This is where the the particle's worldline punctures the boundary graph, for example.  The field reduces to a scalar field in the instance of a spinless particle.   This field has no symmetries of its own.

\medskip

\begin{figure}[h]
\centering
\psfrag{a}{$g_1$}
\psfrag{b}{$g_2$}
\psfrag{c}{$g_3$}
\psfrag{d}{$\vec{P}$}
\psfrag{s}{$s$}
\includegraphics[width = 5cm]{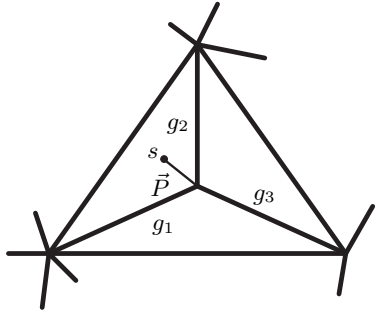}
\caption{Portion of matter coupled to gravity state}
\label{new_particlebound}
\end{figure}

\paragraph{Boundary states} We construct a boundary state.  We illustrate this on the familiar graph with an extra open end for the particle, Figure \ref{new_particlebound}.
The boundary label appearing in the spin foam setting is 
\be
\label{new_spinnetmatter}
D^{j_1}_{m_1n_1}(g_1)\,  D^{j_2}_{m_2n_2}(g_2)\,  D^{j_3}_{m_3n_3}(g_3)\, D^{I}_{mn}\left(S\left(\frac{\vec{P}(g_4g_2g_1^{-1})}{\sin\, m}\right)\right)\, 
 C^{j_1\,j_2\,j_3\, I\,\gLa}_{n_1\,n_2\,n_3\,l},
\ee
where $\gLa$ labels an element in the basis of 4-valent intertwiners.  From the $I = k+s$ representation of the particle sector, we see that when the particle meets the boundary, it has non-zero orbital angular momentum $k$. The particle's momentum is off-shell, but we can of course consider on-shell particle momentum. In a matter field theory, one would expect derivatives of the field occurring in a boundary state for particles with non-zero orbital angular momentum. In the momentum representation, derivatives transform into momenta. The corresponding GFT observable is by projection
\be
\label{new_gftspinob}
\begin{split}
&\int dg_1\, dg_2\, dg_3\, (\vec{P}^{(k}(g_4g_2g_1^{-1})\psi^{s)})_l(g_1^{-1},g_2,g_4)\, \phi(g_2^{-1},g_3,g_{\cdot})\, \phi(g_3^{-1},g_1,g_{\cdot}) \\
&\phantom{xxxxxxxxxxxxxxxxxxxxxxxxxxxxxxxx}\times D^{j_1}_{m_1n_1}(g_1)\, 
 D^{j_2}_{m_2n_2}(g_2)\,  
 D^{j_3}_{m_3n_3}(g_3)\, 
 D^{I}_{mn}\left(S\left(\frac{\vec{P}(g_4g_2g_1^{-1})}{\sin\, m}\right)\right)\, 
  C^{j_1\,j_2\,j_3\, I\,\gLa}_{n_1\,n_2\,n_3\,l},
\end{split}
\ee
where $(\vec{P}^{(k}(g)\psi^{s)})$ means  $k$ copies of the Lie algebra element corresponding to the variable $g$ acting on the spin-$s$ field, and then totally symmetrized. Now onto the perturbative quantum dynamics, where we split our analysis into two parts: the situation of a particle with zero total angular momentum and a particle with arbitrary angular momentum.  

\medskip

\paragraph{Zero total angular momentum}

We propose the action
\be
\label{new_spinlessmatteraction}
\begin{split}
&\cS(\phi,\psi) = \cS(\phi) + \frac{1}{2} \int \prod_{i} dg_i\,d\gal_i  \psi(\gal_2^{-1}g_1\gal_1, \gal_3^{-1}g_2\gal_2, \gal_1^{-1}g_3\gal_3)\,\frac{\sin^2\,\gth - \sin^2 \, m}{i\cos\,\gth} \,   \psi^*(\gal_2^{-1}g_1\gal_1, \gal_3^{-1}g_2\gal_2, \gal_1^{-1}g_3\gal_3)\\
&\phantom{xxxxxxxxxxxxx}+  \frac{\mu_3}{3!}\int \prod_i dg_i \prod_j d\gal_j\,  \psi(\gal_2^{-1}g_1\gal_1, \gal_3^{-1}g_2\gal_2, \gal_1^{-1}g_3\gal_3)\, 
 \psi(\gal_1^{-1}g_5\gal_4, \gal_3^{-1}g_3^{-1}\gal_1, \gal_4^{-1}g_4\gal_3)\\
 &\phantom{xxxxxxxxxxxxxxxxxxxx}\times
 \psi(\gal_{4}^{-1}g_6\gal_2^{-1}, \gal_3^{-1}g_4^{-1}\gal_4, \gal_2^{-1}g^{-1}_2\gal_3) \,
  \phi( \gal_4^{-1}g_5^{-1}\gal_1, \gal_2^{-1}g_6^{-1}\gal_4, \gal_1^{-1}g_1^{-1}\gal_2).\end{split}
\ee
where we have included only a trivalent matter interaction. Bivalent and 4-valent interactions are possible if we include $\psi^2\phi^2$ and $\psi^4$ terms. When we evaluate the related transition amplitudes, we will generate pure gravity amplitudes, amplitudes with particles just propagating, and amplitudes with trivalent particle interactions. It is interesting to note that the action is in two parts: a pure gravity sector and a matter coupled to gravity sector. This accentuates the distinction between these two branches occurring in the causal spin foam models with matter coupling \cite{OriTla_causal}.  The matter cannot propagate on-shell unless the nearby gravity variables do also.  But we can transfer unambiguously to on-shell configurations for the pure gravity sector, as we will show presently.

Now, however, we are more interested in seeing if the group field theory reduces to the effective field theory (EFT) for matter. Recall that the EFT generates a sum over Feynman graphs such that the 3-manifold is a 3-sphere $\cS^3$.  Our group field theory generates also a sum over topologies, so the most we may hope for is that by constraining the variables we might see the EFT as some subset of diagrams.

We recount in brief, the general procedure, for achieving the effective Feynman graphs from a spin foam, before commencing with the analogous procedure in the GFT setting.  In the spin foam setting, we restrict the $3$-topology to be trivial.  But the causal amplitude has two undesirable properties: it is not a gauge-fixed amplitude, and the pure sector is not a topological invariant.  To simplify matters we make the following approximation.  For edges of the cell complex not in the particle graph, we replace the causal amplitude by the a-causal one.  This means the pure gravity sector of the amplitude is a topological invariant.  Thus, we can express it on the simplest possible cell complex, which is the one in which the edges of the particle graph form the totality of the edges of the complex.  Then, we re-express this amplitude so that it may be seen as built from propagators and vertex operators of a field theory.

Now for the group field theory scenario.  At the moment, we cannot systematically gauge-fix, but we wish to deal with the pure gravity sector as above.  To accomplish this approximation we discard the pure gravity terms in the action, and in the trivalent matter term we replace $\phi(g_5^{-1},g_6^{-1},g_1^{-1})$ by  $\gde(g_5^{-1}g_1^{-1}g_6^{-1})$.  This amounts to going on-shell in the pure gravity sector, by replacing $\phi$ with a solution to the classical equation of motion for pure gravity GFT. Therefore, we now have the group field theory
\be
\label{new_reduced}
\begin{split}
&\cS(\psi) =\frac{1}{2} \int \prod_{i} dg_i\,d\gal_i\,  \psi(\gal_2^{-1}g_1\gal_1, \gal_3^{-1}g_2\gal_2, \gal_1^{-1}g_3\gal_3)\,\frac{\sin^2\,\gth - \sin^2 \, m}{i\cos\,\gth} \,   \psi^*(\gal_2^{-1}g_1\gal_1, \gal_3^{-1}g_2\gal_2, \gal_1^{-1}g_3\gal_3)\\
&\phantom{xxxxxxxxxxxxx}+  \frac{\mu_3}{3!}\int \prod_i dg_i \prod_j d\gal_j\,  \psi(\gal_2^{-1}g_1\gal_1, \gal_3^{-1}g_2\gal_2, \gal_1^{-1}g_3\gal_3)\, 
 \psi(\gal_1^{-1}g_5\gal_4, \gal_3^{-1}g_3^{-1}\gal_1, \gal_4^{-1}g_4\gal_3)\\
 &\phantom{xxxxxxxxxxxxxxxxxxxx}\times
 \psi(\gal_{4}^{-1}g_6\gal_2^{-1}, \gal_3^{-1}g_4^{-1}\gal_4, \gal_2^{-1}g^{-1}_2\gal_3) \,\gde(g_5^{-1}g_1^{-1}g_6^{-1}) .
\end{split}
\ee
Now, we use  the Lorentz gauge symmetry occurring in (\ref{new_reduced}) to redefine: $g_3^{-1}\gal_1\rar \gal_1$, $g_2\gal_2\rar\gal_2$, $g_4^{-1}\gal_4\rar\gal_4$ to get
\be
\label{new_reduced2}
\begin{split}
&\cS(\psi) =\frac{1}{2} \int \prod_{i} dg_i\,\gal_i\,  \psi(\gal_2^{-1}g_2g_1g_3\gal_1, \gal_3^{-1}\gal_2, \gal_1^{-1}\gal_3)\,\frac{\sin^2\,\gth - \sin^2 \, m}{i\cos\,\gth} \,   \psi^*(\gal_2^{-1}g_2g_1g_3\gal_1, \gal_3^{-1}\gal_2, \gal_1^{-1}\gal_3)\\
&\phantom{xxxxxxxxxxxxx}+  \frac{\mu_3}{3!}\int \prod_i dg_i \prod_j d\gal_j\, 
 \psi(\gal_2^{-1}g_2g_1g_3\gal_1, \gal_3^{-1}\gal_2, \gal_1^{-1}\gal_3)\, 
 \psi(\gal_1^{-1}g_3^{-1}g_5g_4\gal_4, \gal_3^{-1}\gal_1, \gal_4^{-1}\gal_3)\\
 &\phantom{xxxxxxxxxxxxxxxxxxxx}\times
 \psi(\gal_{4}^{-1}g_4^{-1}g_6g_2^{-1}\gal_2^{-1}, \gal_3^{-1}\gal_4, \gal_2^{-1}\gal_3) \,\gde(g_5^{-1}g_1^{-1}g_6^{-1})\end{split}
\ee
Now, we choose configurations where we can fix the Lorentz symmetry to get\footnote{Gauge-fixing the Lorentz symmetry in the causal scenario still gives a Fadeev-Popov contribution of 1. It is the translation symmetry which is affected by the causality restriction.}
\be
\label{new_reduced3}
\begin{split}
&\cS(\psi) =\frac{1}{2} \int \prod_{i} dg_i\,  \psi(g_2g_1g_3, 1, 1)\,\frac{\sin^2\,\gth - \sin^2 \, m}{i\cos\,\gth} \,  
 \psi^*(g_2g_1g_3, 1, 1)\\
&\phantom{xxxxxxxxxxxxx}+  \frac{\mu_3}{3!}\int \prod_i dg_i \, 
 \psi(g_2g_1g_3, 1, 1)\, 
 \psi(g_3^{-1}g_5g_4, 1, 1)\\
 &\phantom{xxxxxxxxxxxxxxxxxxxx}\times
 \psi(g_4^{-1}g_6g_2^{-1}, 1,1) \,\gde(g_5^{-1}g_1^{-1}g_6^{-1})\end{split}
\ee
And now redefining the variables $g_a = g_2g_1g_3$, $g_b = g_3^{-1}g_5g_4$ and $g_c = g_4^{-1}g_6g_2^{-1}$, and the field $\tilde{\psi}(g) = \psi(g, 1,1)$ gives us the action
\be
\label{new_eftaction}
\cS(\tilde{\psi}) = \frac{1}{2} \int dg\, \tilde{\psi}(g)\,\frac{\sin^2\,\gth - \sin^2 \, m}{i\cos\,\gth} \,  \tilde{\psi}(g^{-1}) +  \frac{\mu_3}{3!}\int dg_a\,dg_b\,dg_c\,  \tilde{\psi}(g_a)\, \tilde{\psi}(g_b)\, \tilde{\psi}(g_c) \,\gde(g_a\,g_b\,g_c)
\ee
which after substituting $ \vec{P} = \vec{n}\,\sin\,\gka\gth$ is the EFT we were looking for.

\medskip

\paragraph{Non-zero total angular momentum}
The action takes the more general form
\be
\label{new_matteraction}
\begin{split}
&\cS(\phi,\psi) = \cS(\phi) + \frac{1}{2} \int \prod_{i} dg_i\,  
\psi^s_{l_1}(\gal_2^{-1}g_1\gal_1, \gal_3^{-1}g_2\gal_2, \gal_1^{-1}g_3\gal_3)\,\cK^s_{l_1l_2}(g_2g_1g_3) \,   (\psi^s)_{l_2}^*(\gal_2^{-1}g_1\gal_1, \gal_3^{-1}g_2\gal_2, \gal_1^{-1}g_3\gal_3)\\
&\phantom{xxxxxxxxxxxxx}+  \frac{\mu_3}{3!}\int \prod_i dg_i \prod_j d\gal_j\,  
(\vec{P}^{(k_a}(g_2g_1g_3)\psi^{s)})_{l_a}(\gal_2^{-1}g_1\gal_1, \gal_3^{-1}g_2\gal_2, \gal_1^{-1}g_3\gal_3)\\ 
&\phantom{xxxxxxxxxxxxxxxxxxxxxxxxxxxxxxxx} \times (\vec{P}^{(k_b}(g_3^{-1}g_5)g_4\psi^{s)})_{l_b}(\gal_1^{-1}g_5\gal_4, \gal_3^{-1}g_3^{-1}\gal_1, \gal_4^{-1}g_4\gal_3)\\
&\phantom{xxxxxxxxxxxxxxxxxxxxxxxxxxxxxxxxxx}\times
 (\vec{P}^{(k_c}(g_4^{-1}g_6g_2^{-1})\psi^{s)})_{l_c}\psi(\gal_{4}^{-1}g_6\gal_2^{-1}, \gal_3^{-1}g_4^{-1}\gal_4, \gal_2^{-1}g^{-1}_2\gal_3)\\
&\phantom{xxxxxxxxxxxxxxxxxxxxxxxxxxxxxxxxxxxx}\times  \phi( \gal_4^{-1}g_5^{-1}\gal_1, \gal_2^{-1}g_6^{-1}\gal_4, \gal_1^{-1}g_1^{-1}\gal_2).\end{split}
\ee
where $K^s_{l_1l_2}(g_3g_2g_1)$ is the same operator as that occurring (\ref{causal_spinningeft}). There are elements $g$ in the spin projector part of the amplitude, these correspond to the framing of the particle graph.   We can, by a similar modification of the action, we may consider a bivalent interaction, and a 4-valent interaction.  The same procedure as for the spinless case serves to give us the effective field theory occurring in (\ref{causal_spinningeft}).

\medskip

\paragraph{Generalizations}

A generalization is needed in the realm of matter coupling.  We are restricted by our action to consider amplitudes with only 4-valent matter interactions.  In this case the generalization is easy. We add interaction terms corresponding to more general convex subsets than the tetrahedron.  For example, one could choose the octahedron, which has eight faces , and so would allow for 8-valent matter interaction. Should we wish, we could propose even more general interaction terms, which have non-triangular faces.  We would then need to define a new fields for the kinematical sector and new kinetic terms to describe their propagation.

\section{Conclusion and outlook}
\label{conclusion}

We wish first to recapitulate what we have accomplished in the paper.  There are two ways to encode the dynamics of point particle motion into a quantum field theory  There is the causal variety with the inverse of the Feynman propagator occurring in the kinetic term and the field interaction in the vertex term.  Also, there are the a-causal models with trivial kinetic operators and the dynamics in the vertex term.  Furthermore, we may consider this a-causal theory as the static ultra-local limit of an underlying more fundamental theory, which is causal.

Then we passed over to the group field theory where we showed that these options are mirrored for pure gravity and gravity coupled to matter. GFTs that have occurred in the literature to date are of the latter variety along with their causal generalization \cite{Ori_gengft}. The generalized case, however, includes ample extra structure which needs interpretation.  This impelled us to seek out a group field theory of the former variety in the hope that we might make contact with field theory as we do in the case of matter.  We found that we should consider the triangulation as the spin foam while the dual cell-complex is realized as the discretization.  This is a fundamental change of view from the original models.  The resulting field theory indeed contains the causal version of the propagator occurring the spin foam models.  Thus, its corresponding Hadamard function is the local simplicial amplitude which when playing in concert with those from all the simplices of a $3$-manifold provides the projection onto the physical subspace of solutions to the Hamiltonian constraint.  This is intrinsically related to the classical solutions of this group field theory where we could expect on-shell configurations.  We clearly saw the most simple solution drop out of the free field equations, and it was indeed a flat geometry. Lorentz symmetry became manifest in the GFT action, which allowed for manipulation of the GFT action.   The remnant of translation symmetry was still obscure, but that should clear when we are able to give a better account of the gauge fixing of the propagator. 

We coupled the group field theory to matter and saw it reduced to the effective field theory for matter occurring in \cite{FreLiv_PR3, OriTla_causal}.  This concluded a lengthy search for such a theory, which seemed not to fall naturally out of earlier attempts to couple matter.  Indeed, we demonstrated that this was a misguided aspiration; since the earlier attempts were in the Hadamard formulation and thus reduce to a different effective field theory.

Thus, we are led to the following assertion.  If we believe in the fundamentality of the group field theory approach, and that it should occur in a form familiar from quantum field theory, then the theory we proposed fits the bill. Of course, all GFT theories differ from usual QFT in the sense that the arguments of the vertex term have a tetrahedral symmetry, and this new GFT is no different. Moreover, there are alternatives, such as those proposed in \cite{Ori_gengft}, which provides the scope to impose causality as a very fundamental level: in terms of explicit orientation labels.  It would be interesting to investigate whether this is possible in this new model, and to see what causal propagator arises at some \lq effective' level.  Also, the earlier group field theories define in a clear fashion an inner product on the state space of quantum gravity, so we need to investigate how this arises in this group field theory.

Our theory offers a chance for much further interesting investigation.  The most pressing generalization is to develop the corresponding theory in 4d. We think that this formalism is intrinsically related to the description of causal states-sum models based on the structure of  a Lie 2-group .  This springs from the simple analysis that the group field theory field in $4d$ represents a tetrahedron and the group arguments the four faces. Thus, if these group elements are to represent holonomies, then they are most naturally the 2-holonomies along the faces.  We could easily write down the corresponding direct generalization of the action here but we shall present the full analysis of the physical meaning of that model in \cite{Rya_proposal2}, and what relation its bears with 4d quantum gravity.

There is also the subject of gauge matter coupling.  Conventionally, matter has been coupled to quantum gravity models at the level of the spin foam amplitudes.  We are caught however in a dilemma.  On the one hand, point matter is coupled as a Feynman diagram.  On the other hand, since both the Yang Mills and gravity theories may be defined non-pertubatively on a lattice, the existing models couple these two theories in that fashion \cite{OriPfe_yangmills}.  Now, the group field theory fits nicely in with the Feynman graph interpretation of matter sources. And this is where our dilemma reaches it crux.  We can either choose to take the view that we should treat the gauge theories on the same level as the matter fields, so that a Feynman graph of the GFT contains a Feynman graph of the gauge theory, matter fields etc.  defined in the spin foam formulation.  Or that the gauge theory should rank in the same fashion as gravity and a given Feynman graph of the GFT should correspond to a non-perturbative formulation of quantum gravity and gauge theory on a given topology, while the matter fields occur as Feynman graphs.

Of course, there has been a recent work done on coupling fermionic fields to quantum gravity at the spin foam level \cite{Fai_fermions}.  Thus, the truth of the matter might be that everything is non-perturbative at the spin foam level, and that the gft is simpliy required to provide a sum over topologies.  But we prefer to deal with the group field theory and take the opposite point of view.  This expectation is heightened by the fact that both the causal amplitudes presented in \cite{OriTla_causal} and the field coupling in \cite{Fai_fermions} are triangulation dependent, so we need a sum over triangulations to retrieve background-independent results.

Finally, there are many aspects of any group field theory that need to be addressed.  First of all, there is the gauge symmetries that characterize the model, at least at the level of the Feynman amplitudes.  These must be gauge-fixed in order to make sensible predictions.  Then we must perform a complete analysis of the non-perturbative aspects of the group field theory, including a calculation of the instantons which should provide a link to the classical regime.  The classical regime of the field theory is very important since it is essentially the physical inner product on a manifold of trivial topology $\gSi\times\R$. There is also the subject of graviton propagation which has been dealt with just recently in \cite{Graviton}.  Starting from a background independent formulation, they derive the graviton propagator by imposing suitable restrictions.  The key idea is to construct n-point functions by means of a propagation kernel, which provides an amplitude for the fields assigned on a boundary of space-time.  They use the group field theory to bolster their claim that a single tetrahedron which is the $O(\lambda)$ contribution will be the leading contribution to the path integral. But this claim is without foundation unless one actually understands what this order-by-order perturbation means.  Furthermore, this is the only use made of the GFT formalism,  the calculation does not use GFT observables thereafter, but loop quantum gravity observables.  To calculate the graviton propagator directly from the group field would provide a validation of this result coming from the use of one coherent formalism.  These issues will be dealt with in future work.

\acknowledgements
I would like to that D. Oriti, T. Tlas and R. Williams for many informative discussions during the course of the project.

\appendix

\section{The bianchi Identity}
\label{app_bianchi}

We present here a specific example.  Consider a vertex $v$ of the discretization $\dman$ such that it has four incident edges $e$.  these edges are dual to faces $\fs$ which form the boundary of a 3-cell in $\dmans$.
\begin{figure}[h]
\centering
\psfrag{a}{$g_1$}
\psfrag{b}{$g_2$}
\psfrag{c}{$g_3$}
\psfrag{d}{$g_4$}
\psfrag{e}{$g_5$}
\psfrag{f}{$g_6$}
\includegraphics[width =5cm]{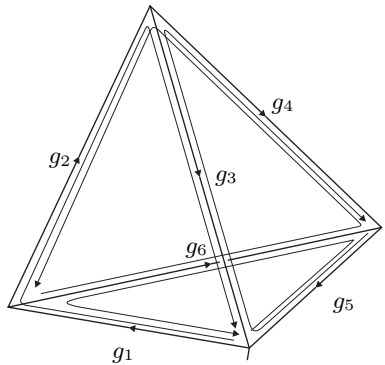}
\caption{The Bianchi Identity}
\label{default}
\end{figure}

There is a parallel transport $g_{\es}$ assigned to each of the edges $\es$:  $g_1,\,g_2,\,g_3,\,g_4,\,g_5,\,g_6$. And the holonomies for the faces are defined to 
\be
\label{bianchi_hol}
G_a = g_2\, g_1\, g_3, \quad\quad G_b = g_3^{-1}\, g_5\, g_4, \quad\quad G_c = g_4^{-1}\, g_6\, g_2^{-1}, \quad\quad G_d = g_6^{-1}\, g_5^{-1}\, g_1^{-1}
\ee
Then, to construct the Bianchi identity one picks an arbitrary base vertex $\vs$ somewhere on the dual ball. In our case we choose the $\vs$ to be the top vertex in the diagram.  The holonomies clearly satisfy the identity 
\be
\label{bianchi_identity}
G_a\, G_b\, G_c\, g_2\, G_d g_2 ^{-1}= 1
\ee
So the specific element $g_{\vs(e)}$, arising in (\ref{bianchi}), is the parallel transport from the chosen base vertex to where the holonomy $G_e$ starts and ends.  From our diagram we can see that only the holonomy $G_d$ does not start and end at the top vertex.  Hence, $g_{\vs{d}} = g_2^{-1}$.



\begin{thebibliography}{11}

\bibitem{Ori_review}
D.~Oriti,
  Rept.\ Prog.\ Phys.\  {\bf 64} (2001) 1489
  [arXiv:gr-qc/0106091]

\bibitem{Per_review}
A Perez,
  [arXiv:gr-qc/0409061]
  
\bibitem{Ash_lectures} 
A  Ashtekar, {\it Lectures on Non-Perturbative Canonical Quantum Gravity},  World Scientific (1991) 

\bibitem{RovSmo_loop} 
C Rovelli, L Smolin, Phys. Rev. Lett. 61, (1988) 1155

\bibitem{Topological}
G Ponzano, T Regge, in {\it Spectroscopy and group theoretical methods in physics}, ed. F Block, North Holland (1968)\\
V Turaev, O Viro, Topology 31, (1992) 865\\
L Crane, D Yetter, in {\it Quantum Topology}, eds. L Kauffman, R Baadhio, World scientific, (1993) 120, [arXiv:hep-th/9301062]\\
L Crane, L Kauffman, D Yetter, J. Knot Theor. Ramifications 6, (1997) 177, [arXiv:hep-th/9409167]\\
J W Barrett, B W Westbury, Trans. Am. Math. Soc. 348, (1996) 3997, [arXiv:hep-th/9311155]

\bibitem{Ori_gftreview}
D Oriti, 
 in {\it Mathematical and Physical Aspects of Quantum Gravity}, 
 B  Fauser, J  Tolksdorf and E  Zeidler, eds., Birkh\"auser, Basel (2006).  [arXiv:gr-qc/0512103]\\
D Oriti,
  [arXiv:gr-qc/0607032]

\bibitem{Fre_gftreview}
L Freidel,
  Int.\ J.\ Theor.\ Phys.\  {\bf 44} (2005) 1769,
  [arXiv:hep-th/0505016]

\bibitem{NouPer_lqgphysical}
K Noui, A Perez,
  Class.\ Quant.\ Grav.\  {\bf 22} (2005) 1739,
  [arXiv:gr-qc/0402110]
  
\bibitem{tHo_polygon}
G 't Hooft,
  Class.\ Quant.\ Grav.\  {\bf 10} (1993) 1653,
  [arXiv:gr-qc/9305008]
  
\bibitem{NouPer_lqqparticles}
 K Noui, A Perez,
  Class.\ Quant.\ Grav.\  {\bf 22} (2005) 4489,
  [arXiv:gr-qc/0402111]

\bibitem{FreLou_PR1}
L Freidel, D Louapre,
  Class.\ Quant.\ Grav.\  {\bf 21} (2004) 5685,
  [arXiv:hep-th/0401076]

\bibitem{FreLou_PR2}
L Freidel,D Louapre,
  [arXiv:gr-qc/0410141]
  
\bibitem{FreLiv_PR3}
L Freidel, E R Livine,
  Class.\ Quant.\ Grav.\  {\bf 23} (2006) 2021,
  [arXiv:hep-th/0502106]
  
\bibitem{Wei_qft}
S Weinberg, {\it The Quantum Theory of Fields, Vol. 1: Foundations}, CUP (1995)


\bibitem{KarLivOriRya_spinning}
M Karadi, E Livine, D Oriti, J Ryan, Effective non-commutative field theory for spinning particles coupled to 3d quantum gravity (In preparation.)


\bibitem{FreOriRya_gftscalar}
L Freidel, D Oriti, J Ryan,
  [arXiv:gr-qc/0506067]

\bibitem{OriRya_gftspinning}
D Oriti,J Ryan,
  Class.\ Quant.\ Grav.\  {\bf 23} (2006) 6543,
  [arXiv:gr-qc/0602010]

\bibitem{LivOri_causal}
E R Livine, D Oriti,
  Nucl.\ Phys.\ B {\bf 663} (2003) 231,
  [arXiv:gr-qc/0210064]

\bibitem{OriTla_causal}
D Oriti and T Tlas,
  [arXiv:gr-qc/0608116]

\bibitem{Tei_qmechanics}
 C Teitelboim,
  Phys.\ Rev.\ D {\bf 25} (1982) 3159.

\bibitem{Ori_feynman}
D Oriti,
  Phys.\ Rev.\ Lett.\  {\bf 94} (2005) 111301,
  [arXiv:gr-qc/0410134]

\bibitem{Ori_gengft}
D Oriti,
  Phys.\ Rev.\ D {\bf 73} (2006) 061502,
  [arXiv:gr-qc/0512069]
  
\bibitem{Superspace} 
 S Giddings, A Strominger, Nucl. Phys. B {\bf 321}, (1989) 481\\
 T Banks, Nucl. Phys. B {\bf 309}, (1988) 493\\
 M McGuigan, Phys. Rev. D {\bf 38}, (1988) 3031

\bibitem{Rov_projector}
C.~Rovelli,
  Phys.\ Rev.\ D {\bf 59} (1999) 104015,
  [arXiv:gr-qc/9806121]

\bibitem{FreLou_diffeo}
L Freidel, D Louapre,
  Nucl.\ Phys.\ B {\bf 662} (2003) 279,
  [arXiv:gr-qc/0212001]

\bibitem{FreLiv_noncompact}
 L Freidel, E R Livine,
  J.\ Math.\ Phys.\  {\bf 44} (2003) 1322,
  [arXiv:hep-th/0205268]

\bibitem{Matrix}
F David, Nucl. Phys. {\bf B}257, (1985) 45\\
J Ambjorn, B Durhuus, J Frolich, Nucl. Phys. {\bf B}257, (1985) 433\\
V A Kazakov, I K Kostov, A A Migdal, Phys. Lett. 157, (1985) 295. \\
D V Boulatov, V A Kazakov, I K Kostov, A A Migdal, Nucl. Phys. {\bf B}275, (1986) 641\\
M Douglas, S Shenker, Nucl. Phys. {\bf B}335, (1990) 635\\
D Gross, A Migdal, Phys. Rev. Lett. 64, (1990) 635\\
E Brezin, V A Kazakov, Phys. Lett. {\bf B}236, (1990) 144

\bibitem{Bou_gft}
D V Boulatov,
  Mod.\ Phys.\ Lett.\ A {\bf 7} (1992) 1629,
  [arXiv:hep-th/9202074]
  
\bibitem{GirOecPer_topological}
 F Girelli, R Oeckl, A Perez,
  Class.\ Quant.\ Grav.\  {\bf 19} (2002) 1093,
  [arXiv:gr-qc/0111022]

\bibitem{Pfe_gauge} 
 H Pfeiffer,
  Annals Phys.\  {\bf 308} (2003) 447,
  [arXiv:hep-th/0304074]

\bibitem{DepPet_gft}
R DePietri, C Petronio,
  J.\ Math.\ Phys.\  {\bf 41} (2000) 6671,
  [arXiv:gr-qc/0004045]

\bibitem{Rya_proposal2}
J Ryan, A new proposal for group field theory II: the 4d case

\bibitem{OriPfe_yangmills}
 D Oriti, H Pfeiffer,
  Phys.\ Rev.\ D {\bf 66} (2002) 124010,
  [arXiv:gr-qc/0207041]
  
\bibitem{Fai_fermions}
W J Fairbairn,
  [arXiv:gr-qc/0609040]


\bibitem{Graviton}
C.~Rovelli,
  Phys.\ Rev.\ Lett.\  {\bf 97} (2006) 151301,
  [arXiv:gr-qc/0508124]\\
S Speziale,
  JHEP {\bf 0605} (2006) 039,
  [arXiv:gr-qc/0512102]
F Mattei, C Rovelli, S Speziale, M Testa,
  Nucl.\ Phys.\ B {\bf 739} (2006) 234,
  [arXiv:gr-qc/0508007]\\
E Bianchi, L Modesto, C Rovelli, S Speziale,
  Class.\ Quant.\ Grav.\  {\bf 23} (2006) 6989,
  [arXiv:gr-qc/0604044]

\end{thebibliography}
\end{document}